\documentclass[%
reprint,
superscriptaddress,
preprintnumbers,
showkeys,
aps,
aps,
prd,
]{revtex4-2}

\usepackage{float}
\usepackage{graphicx}
\usepackage{dcolumn}
\usepackage{bm}
\usepackage{bbold}
\usepackage{braket}
\usepackage{amssymb,amsmath}

\usepackage{color}
\usepackage{amsfonts,mathrsfs}
\usepackage{subfigure}
\usepackage{array}
\usepackage{enumerate}

\newcolumntype{L}{>{\centering\arraybackslash}m{3cm}}

\definecolor{blue}{rgb}{0,0,1}

\definecolor{green}{rgb}{0,1,0}

\definecolor{red}{rgb}{1,0,0}

\definecolor{gray}{rgb}{.5,.5,.5}

\definecolor{darkgreen}{rgb}{.0,.5,.0}

\newcommand{\RNum}[1]{\uppercase\expandafter{\romannumeral #1\relax}}

\usepackage{xspace}
\usepackage{siunitx}
\usepackage{xfrac}
\usepackage{hyperref}
\usepackage[nameinlink]{cleveref}
\usepackage{appendix}
\usepackage{xifthen}
\usepackage{xcolor}
\hypersetup{
	colorlinks,
	linkcolor={red!75!black},
	citecolor={blue!75!black},
	urlcolor={blue!75!black}
}

\graphicspath{{./figures/}{./}}

\begin{document}
	
\title{The dynamical holographic QCD method for hadron physics and QCD matter}

\author{Yidian Chen}
\email[]{chenyidian@ucas.ac.cn} 
\affiliation{School of Nuclear Science and Technology, University of Chinese Academy of Sciences, Beijing 100049, China} 

\author{Danning Li}
\email[]{lidanning@jnu.edu.cn}
\affiliation{Department of Physics and Siyuan Laboratory, Jinan University, Guangzhou 510632, P.R. China} 

\author{Mei Huang}
\email[]{huangmei@ucas.ac.cn}
\affiliation{School of Nuclear Science and Technology, University of Chinese Academy of Sciences, Beijing 100049, China}

	
\begin{abstract}
In this paper we present a short overview on the dynamical holographic QCD method for hadron physics and QCD matter. The 5-dimensional dynamical holographic QCD model is constructed in the graviton-dilaton-scalar framework with the dilaton background field $\Phi$ and the scalar field $X$ dual to the gluon condensate and the chiral condensate operator thus can represent the gluodynamics (linear confinement) and chiral dynamics (chiral symmetry breaking), respectively. The dilaton background field and the scalar field are a function of the 5th dimension, which plays the role of the energy scale, 
in this way, the DhQCD model can resemble the renormalization group from ultraviolet (UV) to infrared (IR). By solving the Einstein equation, the metric structure at IR is automatically deformed by the nonperturbative gluon condensation and chiral condensation in the vacuum. We review the results on the hadron spectra including the glueball spectra, the light/heavy meson spectra, as well as on QCD phase transitions, and thermodynamical as well as transport properties in the framework of the dynamical holographic QCD model.

\end{abstract}

\maketitle
\tableofcontents

\section{Introduction}
\label{intro}
Quantum chromodynamics (QCD) is the fundamental theory of strong interaction describing more than 99$\%$ of visible matter in the universe. Though it is quite successful in the ultraviolet region when the coupling is weak\cite{Gross:1973id,Politzer:1973fx}, it remains as an outstanding challenge to solve nonperturbative QCD physics in the infrared (IR) regime, especially the chiral symmetry breaking and color confinement related to hadron physics as well as QCD phase transitions. To tackle this problem, some important non-perturbative methods have been developed, especially lattice QCD \cite{Kogut:1982ds,Gupta:1997nd,Fodor:2012gf,Bloch:2003sk}, Dyson-Schwinger
equations (DSEs)\cite{Alkofer:2000wg,Bashir:2012fs}, and functional renormalization group equations (FRGs)\cite{Wetterich:1992yh,Pawlowski:2005xe,Gies:2006wv}. In recent decades, the anti-de Sitter/conformal field theory (AdS/CFT) correspondence or gauge/gravity duality \cite{Maldacena:1997re,Gubser:1998bc,Witten:1998qj} has been widely used
in dealing with nonperturbative QCD problems in hadron physics and strongly coupled quark matter. 

It is expected that there exists a general holography principle, which maps a D-dimensional quantum field theory (QFT) to a (D + 1)-dimension quantum gravity, and the gravitational description becomes classical when the QFT is strongly-coupled. The extra dimension here can be treated as an emergent energy scale or renormalization group (RG) flow in the QFT \cite{Adams:2012th}. Therefore it is possible to construct a non-conformal 5-dimensional holographic QCD model based on the gauge/gravity duality, and there have been lots of efforts both from top-down and bottom-up approaches. From top-down, the $D_p-D_q$ system including the $D_3-D_7$ \cite{Erdmenger:2007cm} and the $D_4-D_8$ system or the Witten-Sakai-Sugimoto (SS) model \cite{Sakai:2004cn,Sakai:2005yt} have been widely explored. In the bottom-up approach, the hard-wall AdS/QCD model \cite{Erlich:2005qh} and the soft-wall AdS/QCD or KKSS model \cite{Karch:2006pv} established the 5-dimensional framework for light hadron spectra. Interesting progress was made in Refs. \cite{Colangelo:2008us, Ghoroku:2005vt,Gherghetta:2009ac,Sui:2009xe} to incorporate linear trajectories and chiral symmetry breaking. 

The chiral symmetry breaking and the color confinement are two important non-perturbative features of QCD. From the light meson spectra, one can read the information of the chiral symmetry breaking and the linear confinement from the mass difference between the chiral partners and the linear Regge trajectory, respectively. The spontaneous chiral symmetry breaking is well understood by the dimension-3 quark condensate $\langle\bar{q}q\rangle$ in the vacuum and the understanding of color confinement remains a challenge. The linear Regge trajectories of meson spectra suggest that color charges can form the string-like structure inside mesons thus the confinement can manifest itself by the linear potential between quark anti-quark at large distances. The dynamical holographic QCD (DhQCD) model \cite{Li:2012ay,Li:2013oda,Li:2013xpa} has been constructed by promoting the soft-wall model to a dynamical model with gluon dynamics background and the probe of matter part in the graviton-dilaton-scalar framework
, where the dilaton background field $\Phi(z)$ is dual to the gluon operator and the scalar field $X(z)$ is dual to the quark operator at the UV boundary. Evolution of both $\Phi(z)$ and $X(z)$ along the fifth-dimension from ultraviolet (UV) to infrared (IR) resembling the renormalization group. The gluon dynamics background is solved by the coupling between the graviton and the dilaton field $\Phi(z)$ describing the gluon condensate and confinement, and the scalar field mimics chiral dynamics.  The metric structure at IR is automatically deformed by the nonperturbative gluon condensation and chiral condensation in the vacuum, thus the DhQCD model can simultaneously describe both the chiral symmetry breaking and linear confinement.

Except the dynamical holographic QCD model, the Gubser model \cite{Gubser:2008ny,Gubser:2008yx,DeWolfe:2010he} and the improved holographic QCD model \cite{Gursoy:2007cb,Gursoy:2007er,Gursoy:2010fj} with the dilaton potential, the refined model \cite{Yang:2014bqa} and Dudal model \cite{Dudal:2017max} with the deformed metric belong to  the same graviton-dilaton system. In the graviton-dilaton system, the deformed metric, the dilaton field and the dilaton potential should be solved self-consistently from each other through the Einstein equations and the equation of motion of the dilaton field. Therefore, the three types of models, A) with the known form of the dilaton field, B)with the known deformed metric, and C) with the known dilaton potential to solve the other two unknowns of the system are equivalent to describe the background at zero temperature and zero density. 

In the dynamical holographic QCD model, we input the dilaton field $\Phi(z)$ which is dual to the gluon condensate operator at UV. With the deformed metric by the gluon condensation, the dynamical holographic QCD model can well describe the glueball spectra including the scalar glueball \cite{Li:2012ay}, and charge parity even and odd two-gluon and three-gluon glueball spectra \cite{Chen:2015zhh,Zhang:2021itx}.  Considering the backreaction of the scalar field on the background, the dynamical holographic QCD model can describe the light meson spectra and pion form factor\cite{Li:2012ay}. It can also be extended to four flavor case to describe heavy-flavor meson spectra \cite{Chen:2021wzj}. Further studies \cite{Li:2014hja,Li:2014dsa,Chelabi:2015cwn} show that this dynamical holographic QCD model can successfully describe QCD phase transition, thermodynamical properties and transport properties of QCD matter, especially the temperature dependent shear viscosity, bulk viscosity, electric conductivity as well as jet quenching parameter.

    The paper is organized as following: we introduce the general Graviton-dilaton-scalar framework in section \ref{sec-GDS}. Then in section \ref{sec-glueball-meson-spectra} we introduce the glueball spectra, light and heavy flavor meson spectra in the dynamical holographic QCD model. In section \ref{sec-eos-transport} we introduce the thermodynamical and transport properties from the dynamical holographic QCD model, and in section \ref{sec-phasetransition} we introduce how to realize chiral phase transition.  Finally, a short summary is given in section \ref{sec-summary}.

\section{The general graviton-dilaton-Scalar system}
\label{sec-GDS}

To apply the holographic method in QCD, one of the main tasks is to break the conformal symmetry in the original AdS/CFT correspondence. A usual way is to consider the coupling between the background gravity and the dilaton field $\Phi$, in which the QCD relevant scales could be introduced. The 5D effective action for the graviton-dilaton coupled system could be derived from 10D supergravity~\cite{Gursoy:2007cb,Gursoy:2007er,Li:2011hp}  and it takes the following form
\begin{equation}\label{action-ED-string}
S_{GD}= \frac{1}{16\pi G_5}\int
 d^5x\sqrt{g_s}e^{-2\Phi}\big(R+4\partial_m\Phi\partial^m\Phi-V_s(\Phi)\big),
\end{equation}
in which the string frame is taken. Here, $g^S_{MN}$ is the metric in string frame, $g_S$ is the determinant of the metric, $R$ is the Ricci scalar, $\Phi$ is the dilaton field, $G_5$ is the 5D Newton's constant and $V_\Phi$ is the dilaton potential. 

It is also not difficult to make a frame transformation to the Einstein frame $g^S_{MN}=e^{-\frac{4}{3}\Phi}g^E_{MN}$, which is more convenient to study thermodynamical quantities. Then, the 5D action in the new frame reads
\begin{equation} \label{action-ED-Einstein}
S_{GD}=\frac{1}{16 \pi G_5} \int d^5 x
\sqrt{-g^E}\left(R-\frac{4}{3}\partial_{\mu}\phi\partial^{\mu}\phi-V_E(\phi)\right),
\end{equation}
with $V_E(\Phi)=e^{\frac{4}{3}\Phi}V_S(\Phi)$ the dilaton potential in Einstein frame.  

It is proposed that the dilaton field is dual to the effective degrees of freedom of gluons~\cite{Csaki:2006ji-1}. It is also closely related to the Regge slope of the hadronic spectra \cite{Karch:2006pv} or the QCD running coupling constants \cite{Gursoy:2007cb,Gursoy:2007er}. Such a 5D system is shown to give good description of the thermodynamics of pure gluon system~\cite{Li:2011hp}. As for the flavor dynamics, the KKSS model, also named as soft-wall model, provides a good start point, by promoting the 4D global symmetry $SU(N_f)_L\times SU(N_f)_R$ to 5D gauge symmetry. The action of KKSS model takes the following form~\cite{Karch:2006pv} 
\begin{eqnarray}\label{action-kkss}
S_{KKSS} &=& - \int d^5x
 \sqrt{g_s} e^{-\Phi} Tr\Big(|DX|^2+V_{X} \nonumber \\
 & &  +\frac{1}{4g_5^2}(F_L^2+F_R^2)\Big).
\end{eqnarray}
Here X is a matrix-valued scalar field $X^{\alpha\beta}$, which is dual to the operator $\bar{q}^\alpha q^\beta$, with $\alpha,\beta$ flavor indexes. $V_X$ is the potential of the scalar field. The left and right hand Gauge fields
$L_{M}$ and $R_{M}$ are dual to the $SU(N_f)_L\times SU(N_f)_R$ currents, with $F_L$ and $F_R$ their field strength respectively. In such a model, one can consider a lot of interesting physics related with chiral symmetry breaking, like the quark mass effect~\cite{Chelabi:2015gpc,Chen:2018msc}, the multiple flavor effect~\cite{Li:2016smq}, the isospin number density effect~\cite{Lv:2018wfq}, the magnetic effect \cite{Rodrigues:2018pep,Li:2016gfn}, and so on. When all the quark masses are degenerate, one usually takes the expectation value of the $X$ filed as a diagonal form, i.e. $\propto \chi I_{N_f}$, with $I_{N_f}$ the $N_f\times N_f$ identity matrix and $\chi$ a background field dual to the chiral condensate.  It is also possible to introduce the $U_B(1)$ symmetry by adding a $U(1)$ gauge field $\mathcal{F}_{MN}$~\cite{Chen:2019rez,Rodrigues:2020ndy}.  

Generally, the glue-dynamics and the flavor dynamics should be coupled together. Thus, to describe the full QCD dynamics, one has to couple the two systems, Eq.\eqref{action-ED-string} and Eq.\eqref{action-kkss} together, then one reaches the full action of the holographic QCD model
\begin{eqnarray}\label{action-full}
S_{HQCD}=S_{GD}+\lambda S_{KKSS}.
\end{eqnarray}
Here we explicitly write a coupling constant $\lambda$, and we assume $\lambda$ tends to zero and the two systems would decouple when $N_f<<N_c$. In such a system, when considering the background field, one has the metric $g_{MN}$, the dilaton filed $\Phi$, the scalar field $\chi$, thus we will call it the graviton-dilaton-scalar system. When considering finite density effect, one can add the  the $U(1)$ field $\mathcal{F}_{MN}$, and consider a general graviton-Maxwell-dilaton-scalar system. By self-consistently solving all the background out of the equations of motion, one can build the dynamical holographic QCD model.

\begin{figure}[!htb]
\includegraphics[scale=.4]{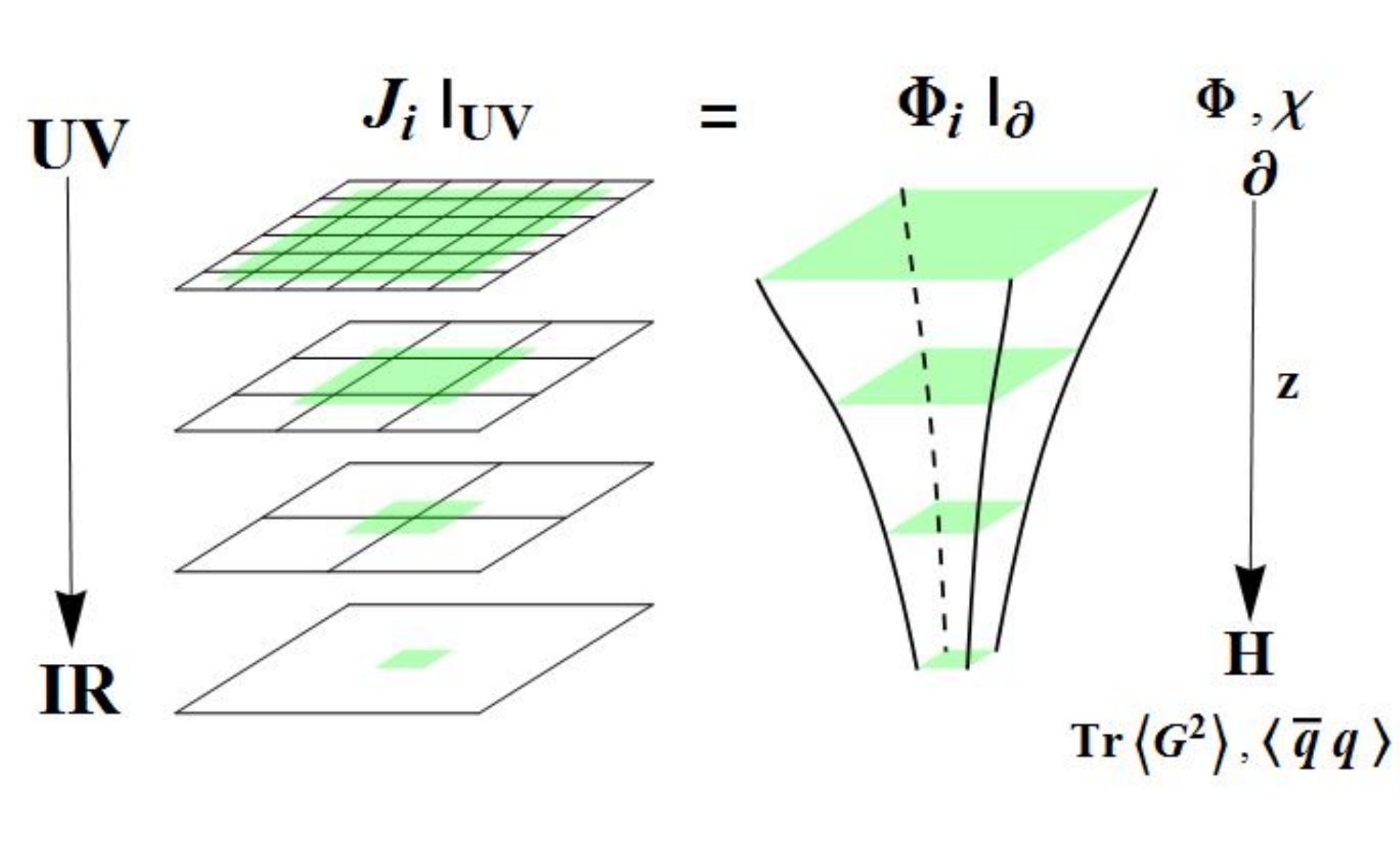}
\caption{A skectch plot for the duality between $d$-dimension QFT and $d+1$-dimension gravity as shown
in \cite{Adams:2012th} (Left-hand side). The RG flow from UV to IR (Right-hand side) are resembled in the dynamical holographic QCD model through the evolution of the fileds in the holographic dimension. The figure is taken from Ref.~\cite{Huang:2013qvz}. }
\label{fig:RGflow}
\end{figure}

In the dynamical holographic model, the dilaton field $\Phi(z)$ and the scalar field $X(z)$ are assumed to be dual to the gluo-dynamics and chiral dynamics, respectively. More concretely, the dilaton field $\Phi(z)$ is dual to a certain kind of gluonic operator(most likely the dimension 4 gluon condensate $G^2$), while the scalar
field $X(z)$ is dual to the chiral condensate $\bar{q}q$. The non-trivial vacuum structure of QCD indicates non-zero vacuum expectation values of those operators, which implies non-trivial boundary conditions of those background fields at UV boundary. Then, from the equation of motion, one can dynamically solve the IR evolution of both the background field and the metric. In this way, the dynamical holographic QCD could resemble the renormalization group from UV to IR with the extra dimension being interpreted as an energy scale, just as pointed out in Ref.\cite{Adams:2012th} and as shown in Fig.\ref{fig:RGflow}. 

In the following sections, we will consider the dynamical holographic QCD model in different cases.

\section{The glueball spectra and meson spectra}
\label{sec-glueball-meson-spectra}

\subsection{DhQCD for Glueball spectra}
In this subsection, the spectra of the glueballs and oddballs are investigated. For studying glueballs/oddballs in the holographic model, there are two separate approaches, one is called "glueball fluctuation approach", which treats scalar, vector and tensor fluctuations on the 5-dimensional background metric as eigenstates of glueballs of various $J^{PC}$. This approach is commonly used in top-down models, as in Ref. \cite{Gursoy:2009jd,Brower:2000rp,Apreda:2003sy,Ballon-Bayona:2017sxa,Elander:2020csd,Gursoy:2007er}. The other approach, the one used in this work, is known as the "glueball excitation method", which treats glueballs as Kaluza-Klein excited states by introducing the effective action of distinct glueballs on the deformed AdS$_5$ spacetime background. In this method, the total action is
\begin{align}
   & S_{\text {total }}=S_{GD}+S_{g},
  \label{action_total}
\end{align}
where $S_{g}$ is the effective action of the glueball/oddball in the string frame.
This approach is widely used in bottom-up models such as Ref. \cite{Colangelo:2007pt,Forkel:2007ru,Bellantuono:2015fia,FolcoCapossoli:2015jnm,FolcoCapossoli:2016fzj}, which treats glueball spectra like other meson spectra. The glueball excitation method was employed in this work to ensure that the meson and glueball spectra are treated in an equal way in the DhQCD model.

The glueballs/oddballs are excited from the pure gluon background at zero temperature and zero chemical potential, i.e. the action Eq.(\ref{action-ED-string}). The background metric is assumed to have the following form
\begin{align}
  ds^{2}=\frac{L^2 e^{2 A_{s}(z)}}{z^{2}}\left(-f(z) d t^{2}+\frac{d z^{2}}{f(z)}+dy_{i}^{2}\right),
  \label{metric_str}
\end{align}
with the $AdS_5$ radius $L$ which is set to 1 and the blackening factor $f(z)$.
Applying the Weyl transformation, the background action can be converted to the Einstein frame Eq.\eqref{action-ED-Einstein}. 
The metrics in Einstein frame are shown as
\begin{align}
  d s^{2}=\frac{L^2 e^{2 A_{E}(z)}}{z^{2}}\left(-f(z) d t^{2}+\frac{d z^{2}}{f(z)}+d y_{i}^{2}\right),
  \label{metric_Einstein}
\end{align}
with $A_E(z)=A_s(z)-\sqrt{\frac{1}{6}}\phi(z)$.
In the Einstein frame, the equation of motion of the background is
\begin{align}
   & f^{\prime \prime}+f^{\prime}\left(-\frac{3}{z}+3 {A_{E}}^{\prime}\right)=0,
  \label{EOMs_2_2}                                                                                                                                                                                                                                                                                     \\
   & A_{E}^{\prime \prime} +\frac{f^{\prime \prime}}{6 f}+A_{E}^{\prime}\left(-\frac{6}{z}+\frac{3 f^{\prime}}{2 f}\right)-\frac{1}{z}\left(-\frac{4}{z}+\frac{3 f^{\prime}}{2 f}\right)\nonumber\\
   & \qquad+3 {A_{E}}^{\prime 2} +\frac{L^2 e^{2 {A_{E}}} V_{\phi}}{3 z^{2} f}=0,
  \label{EOMs_2_3}                                                                                                                                                                                                                                                                                     \\
   & A_{E}^{\prime \prime}-A_{E}^{\prime}\left(-\frac{2}{z}+A_{E}^{\prime}\right)+\frac{\phi^{\prime 2}}{6}=0,
  \label{EOMs_2_4}                                                                                                                                                                                                                                                                                     \\
   & \phi^{\prime \prime}+\phi^{\prime}\left(-\frac{3}{z}+\frac{f^{\prime}}{f}+3 A_{E}^{\prime}\right)-\frac{L^2 e^{2 {A_{E}}}}{z^{2} f} \frac{\mathrm{d} V_{\phi}(\phi)}{\mathrm{d} \phi} =0,
  \label{EOMs_2_5}
\end{align}
with $\phi=\sqrt{\frac83}\Phi$. In the string frame, the effective action of the scalar, vector, and tensor glueballs/oddballs are
\begin{eqnarray}
S_{\mathscr{G}}&=&-\frac{1}{2}\int d^5 x \sqrt{g_s}e^{-p\Phi}(\partial_M \mathscr{G}\partial^M\mathscr{G}\nonumber\\
& & +M_{\mathscr{G},5}^2(z) \mathscr{G}^2) \,  \label{action-SAV-S}\\
S_{V}&=&-\frac{1}{2}\int d^{5}x\sqrt{g_s}e^{-p\Phi}(\frac{1}{2}F^{MN}F_{MN}\nonumber\\
& & +M_{\mathscr{V},5}^2(z) \mathscr{V}^{2}), \label{action-SAV-V}\\
S_{T}&=&-\frac{1}{2}\int d^{5}x\sqrt{g_s}e^{-p\Phi}(\nabla_{L}h_{MN}\nabla^{L}h^{MN}\nonumber\\
&-&2\nabla_{L}h^{LM}\nabla^{N}h_{NM}+2\nabla_{M}h^{MN}\nabla_{N}h \nonumber \\
&-&\nabla_{M}h\nabla^{M}h+M_{h,5}^{2}(z)(h^{MN}h_{MN}-h^{2})),
\label{action-SAV-T}
\end{eqnarray}
where
\begin{eqnarray}
M_5^2(z)=M_5^2 e^{-c_{\rm r.m.}\Phi},
\end{eqnarray}
with constant $c_{\rm r.m.}$ which is used to fit the glueballs spectra. The 5-dimensional mass squared $M_5^2$ is obtained according to the AdS/CFT dictionary, which relates the $q$-form field mass and the conformal dimension $\Delta$ of its dual operator\cite{Maldacena:1997re,Gubser:1998bc,Witten:1998qj}
\begin{eqnarray}
M_5^2=(\Delta-q)(\Delta+q-4).
\end{eqnarray}
The different P-parity glueballs are distinguished by the parameter $p$,  and $p = 1$ for even and $p = -1$ for odd. The gauge invariant operators for different $J^{PC}$ glueballs can be obtained from Ref. \cite{Tang:2015twt,Pimikov:2017bkk,Chen:2021cjr,Brower:2000rp}. For the action of the tensor glueballs, the spin-2 field $h_{MN}$ satisfies the following conditions
\begin{eqnarray}
h=h_M^M=0,\quad \nabla_Mh^{MN}=0.
\end{eqnarray}
For higher spin glueballs(spin $\geq 3$), the action is shown in Ref. \cite{Karch:2006pv,Zhang:2021itx} and will not be shown specifically here.

The equations of motion of the glueballs with different spin can be obtained from the action Eqs.(\ref{action-SAV-S},\ref{action-SAV-V},\ref{action-SAV-T}). The equations of motion of the scalar glueball are given as
\begin{eqnarray}
  -{\mathscr{G}_n}^{''}+V_{\mathscr{G}} {\mathscr{G}_n}=m_{\mathscr{G},n}^2 {\mathscr{G}_n},
  \label{EOM-glueball}
\end{eqnarray}
where the potential in the Schr\"{o}dinger-like equation is
\begin{eqnarray}
  V_{\mathscr{G}}&=&\frac{3A_s^{''}+\frac{3}{z^2}-p\Phi^{''}}{2}+\frac{\left[3A_s^{'}-\frac{3}{z}-p\Phi^{'}\right]^2}{4}\nonumber\\
  && +\frac{1}{z^2}e^{2A_s} \mathrm{e}^{-c_{\text{r.m.}}\Phi} M_{\mathscr{G}, 5}^{2}.
  \label{potential-glueball-s}
\end{eqnarray}
The equations of motion for the vector glueballs have the following form
\begin{eqnarray}
  -\mathscr{V}_n^{''}+V_{\mathscr{V}} \mathscr{V}_n=m_{\mathscr{V},n}^2 \mathscr{V}_n,
\end{eqnarray}
where the potential is written as
\begin{eqnarray}
  V_{\mathscr{V}}&=&\frac{A_s^{''}+\frac{1}{z^2}-p\Phi^{''}}{2}+\frac{\left[A_s^{'}-\frac{1}{z}-p\Phi^{'}\right]^2}{4}\nonumber\\
  &&+\frac{1}{z^2}\mathrm{e}^{2A_s} \mathrm{e}^{-c_{\text{r.m.}}\Phi} M_{\mathscr{V},5}^2.
  \label{potential-glueball-v}
\end{eqnarray}
The Schr\"{o}dinger-like equations for the tensor glueballs are
\begin{eqnarray}
  -\mathscr{T}_n^{''}+V_{\mathscr{T}} \mathscr{T}_n=m_{\mathscr{T},n}^2 \mathscr{T}_n,
\end{eqnarray}
with the potential
\begin{eqnarray}
  V_{\mathscr{T}}&=&\frac{3A_s^{''}+\frac{3}{z^2}-p\Phi^{''}}{2}+\frac{\left[3A_s^{'}-\frac{3}{z}-p\Phi^{'}\right]^2}{4}\nonumber\\
  &&+\frac{1}{z^2}\mathrm{e}^{2A_s} \mathrm{e}^{-c_{\text{r.m.}}\Phi} M_{\mathscr{T}, 5}^{2}.
  \label{potential-glueball-s2_t}
\end{eqnarray}
For higher spin glueballs $\mathscr{H}$ (spin $\geq 3$), the equations of motion have the similar form
\begin{eqnarray}
  -\mathscr{H}_n^{''}+V_{\mathscr{H}} \mathscr{H}_n=m_{\mathscr{H},n}^2 \mathscr{H}_n,
\end{eqnarray}
where the potential in the equation is
\begin{eqnarray}
  V_{\mathscr{H}}&=&\frac{\left(2S-1\right)A_s^{''}+\frac{2S-1}{z^2}-p\Phi^{''}}{2}\nonumber\\
  &&+\frac{\left[\left(2S-1\right)A_s^{'}-\frac{2S-1}{z}-p\Phi^{'}\right]^2}{4}\nonumber\\
  &&+\frac{1}{z^2}\mathrm{e}^{2A_s} \mathrm{e}^{-c_{\text{r.m.}}\Phi} M_{\mathscr{H}, 5}^{2}.
  \label{potential_glueball_high_spin}
\end{eqnarray}

\begin{figure}[h]
\includegraphics[width=0.4\textwidth]{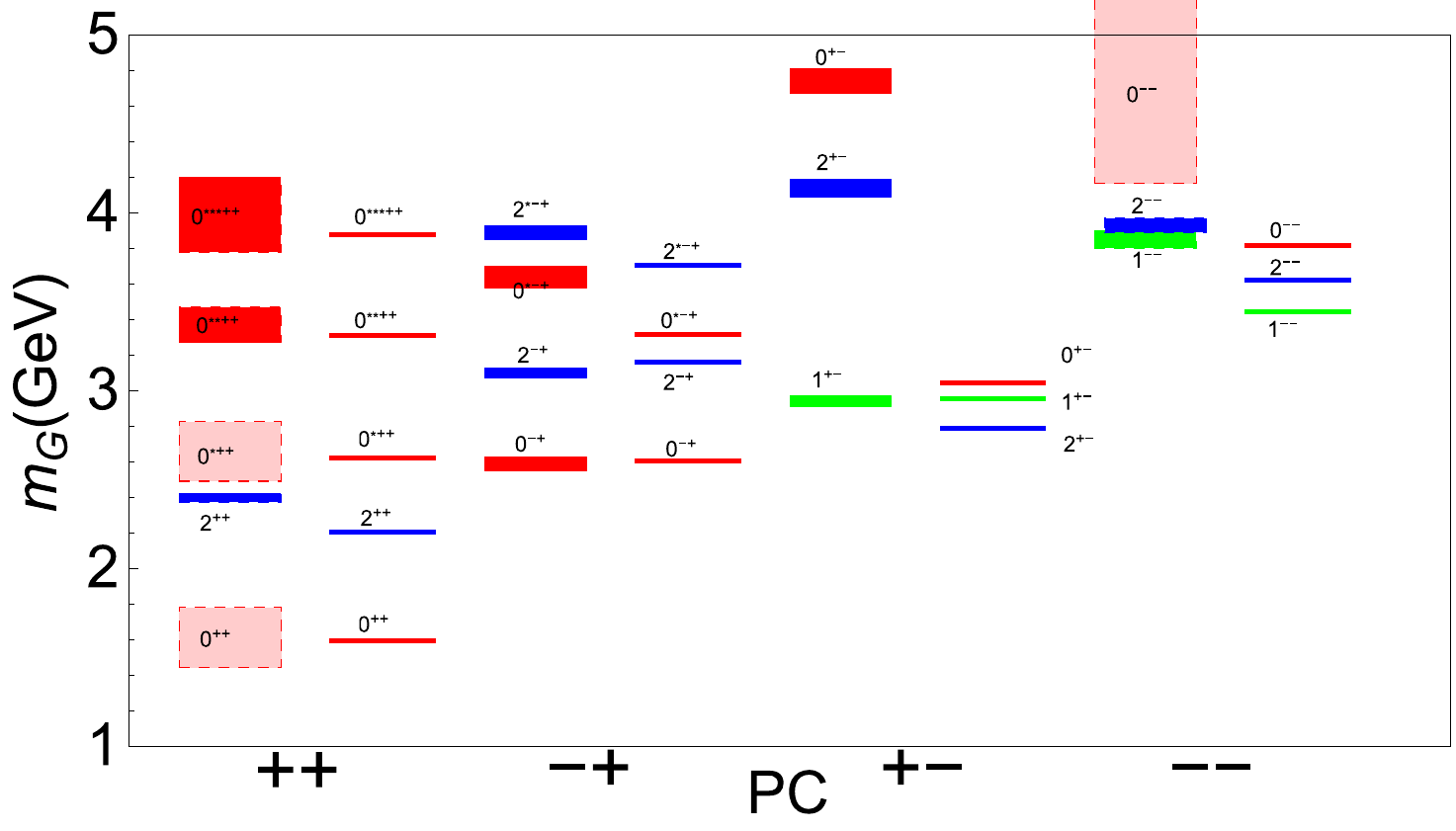} \hspace*{0.1cm}
\caption[]{The glueball/oddball spectra in the DhQCD model compared with the lattice results \cite{Morningstar:1999rf,Lucini:2001ej,Meyer:2004gx,Chen:2005mg,Gregory:2012hu}. (The rectangles in the figure. The figure is from Ref. \cite{Chen:2015zhh}.)} \label{spectra}
\end{figure}

By choosing $c_{\rm r.m.} = 2/3$, the DhQCD model has only one free parameter $b=1$GeV$^2$, which is fixed by the Regge slope of the $0^{++}$ scalar glueball spectra. Here, for solving the graviton-dilaton background, the dilaton configuration $\Phi=b z^2$ is chosen. The final results of the glueball/oddball spectra are shown in Fig.\ref{spectra}, which are in good agreement with the lattice data, except for the three trigluon glueballs $0^{--}$, $0^{+-}$ and $2^{+-}$, which are about 1.5 GeV lighter than the lattice data.

\begin{figure}
  \begin{center}
    \includegraphics[width=0.4\textwidth,clip=true,keepaspectratio=true]{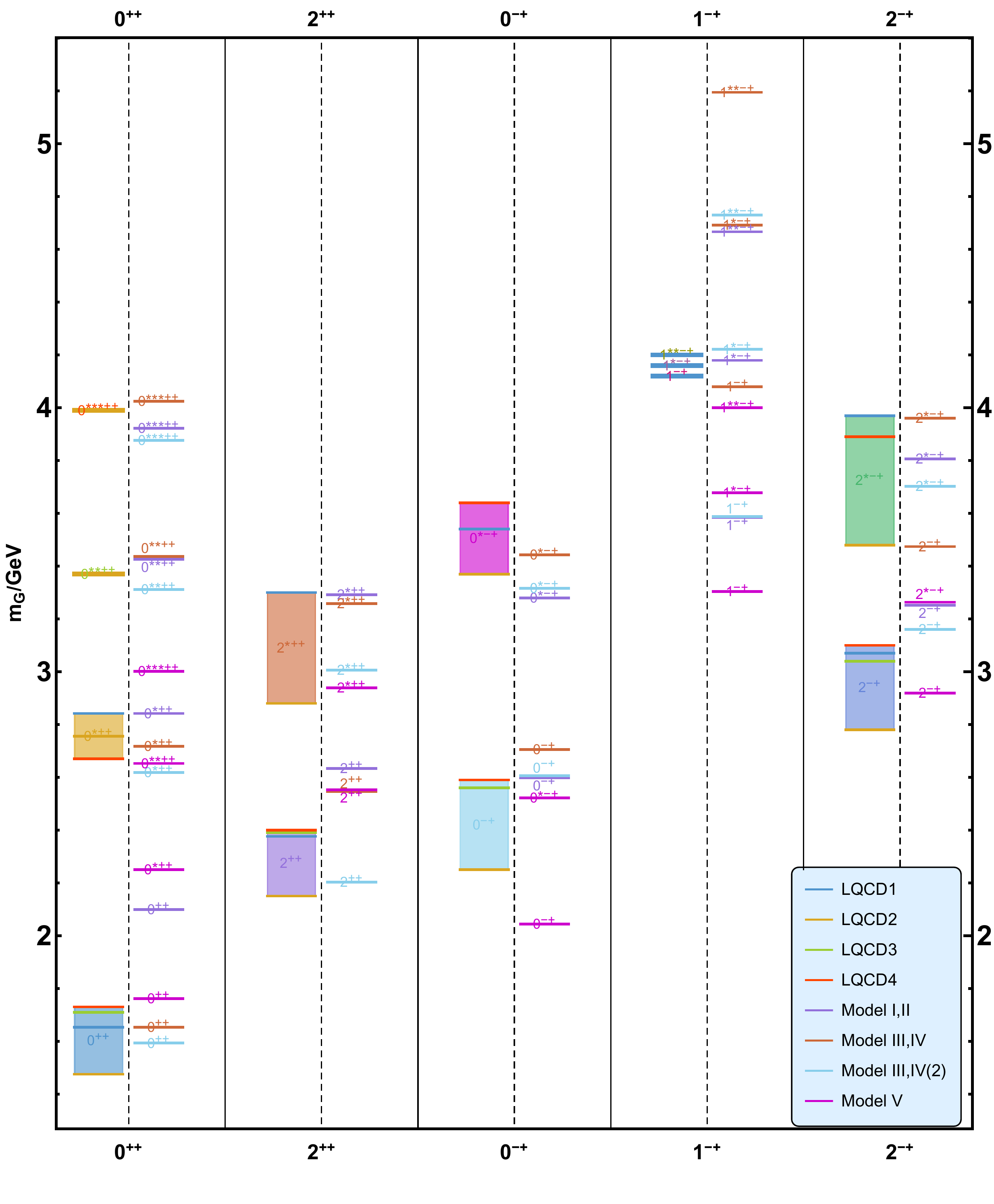}
  \end{center}
  \caption{The spectra of distinct glueballs ($C=1$) in the DhQCD model and compared with lattice results. The horizontal coordinates of the graph indicate the $J^{PC}$ of the glueballs and the vertical coordinates indicate their mass. The rectangles on the figure are the lattice data \cite{Athenodorou:2020ani,Meyer:2004gx,Chen:2005mg,Morningstar:1999rf}, while the solid horizontal line denotes the DhQCD model results. The different color lines represent different models, where medium purple lines, sienna lines, sky blue lines, and magenta lines represent "Model \RNum{1},\RNum{2}", "Model \RNum{3},\RNum{4}(1)", "Model \RNum{3},\RNum{4}(2)", and "Model \RNum{5}", respectively. The figure is from Ref. \cite{Zhang:2021itx}. }
  \label{fig_glueb_spect_holog_and_latti_C_even}
\end{figure}

\begin{figure}
  \begin{center}
    \includegraphics[width=0.4\textwidth,clip=true,keepaspectratio=true]{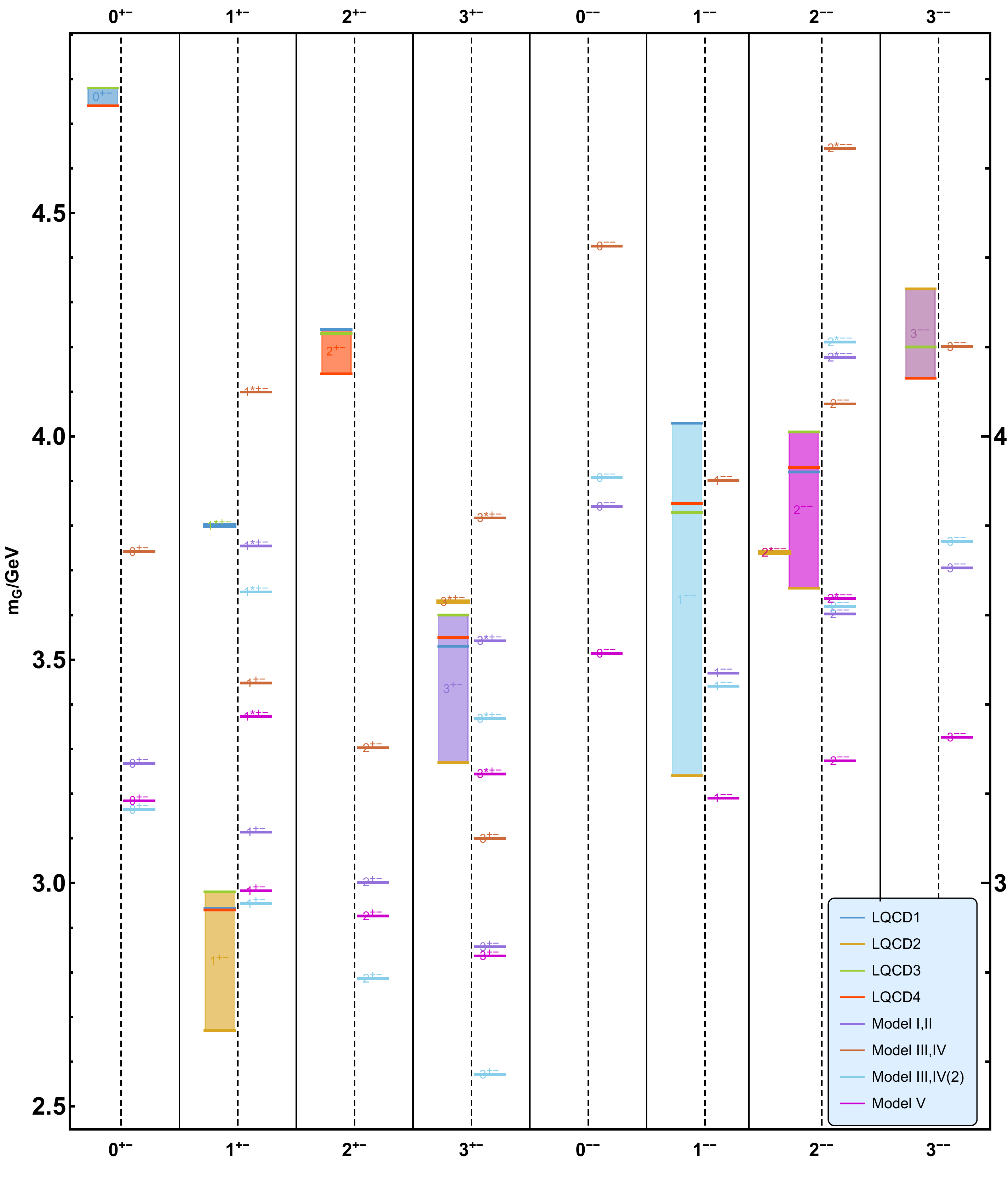}
  \end{center}
  \caption{The spectra of distinct oddballs ($C=-1$) in the DhQCD model and compared with lattice results. The horizontal coordinates of the graph indicate the $J^{PC}$ of the oddballs and the vertical coordinates indicate their mass. The rectangles on the figure are the lattice data \cite{Athenodorou:2020ani,Meyer:2004gx,Chen:2005mg,Morningstar:1999rf}, while the solid horizontal line denotes the DhQCD model results. The different color lines represent different models, where medium purple lines, sienna lines, sky blue lines, and magenta lines represent "Model \RNum{1},\RNum{2}", "Model \RNum{3},\RNum{4}(1)", "Model \RNum{3},\RNum{4}(2)", and "Model \RNum{5}", respectively.The figure is from Ref. \cite{Zhang:2021itx}. }
  \label{fig_glueb_spect_holog_and_latti_C_odd}
\end{figure}

Since there exist two ways to solve the gravtion-dilation system, i.e., choosing the dilaton configuration $\phi=\sqrt{\frac83}\Phi$ and the deformed metric $A_E$, four possible cases are considered. For "Model \RNum{1},\RNum{2}", the deformed metric $A_E=-az^2$ is chosen with $a=0.4822$GeV$^2$ and $c_{\rm r.m.} = 0.4245$ is considered. For "Model \RNum{3},\RNum{4}(1)", the dilaton configuration $\phi=bz^2$ is chosen with $b=1.5360$GeV$^2$ and $c_{\rm r.m.} = 0.4593$ is fixed. To compare with the results of Fig.\ref{spectra}, "Model \RNum{3},\RNum{4}(2)" sets the parameter $c_{\rm r.m.}$ to $2/3$ as before, where the configuration of dilaton is unchanged. Finally, a disparate dilaton configuration $\phi$ is taken into account,
\begin{equation}
\phi(z)=\frac{2 \sqrt{6}}{3} z \sqrt{3 d\left(3+2 d z^{2}\right)}+6 \operatorname{arcsinh}\left[\sqrt{\frac{2 d}{3}} z\right],
\end{equation}
where the parameter $d$ is chosen as 0.2463GeV$^2$ and the parameter $c_{\rm r.m.}$ is fixed as $0.3576$, named as "Model \RNum{5}". The spectra of glueball and oddball for the four cases of DhQCD model are shown in Fig.\ref{fig_glueb_spect_holog_and_latti_C_even} and \ref{fig_glueb_spect_holog_and_latti_C_odd}, respectively.

According to the AdS/CFT dictionary, glueballs with the same spin $J$ and $C$-Parity but different $P$-Parity have the same five-dimensional masses, although they have different operators. To distinguish the glueball states in this case, the model introduces positive and negative dilaton couplings in the action, i.e., $e^{-p\Phi}$. In addition, the model includes two parameters, one in the dilaton configuration or deformed metric $A_E$ and the other in the modified five-dimensional mass squared. With these two parameters, the DhQCD model calculates the different glueball/oddball states, the results of which are compared with the lattice data and shown in Fig.\ref{spectra}, \ref{fig_glueb_spect_holog_and_latti_C_even} and \ref{fig_glueb_spect_holog_and_latti_C_odd}. As can be seen from the figure, the predicted spectra of the DhQCD model for the glueball/oddball are in remarkably good agreement with the lattice results, except for the three oddball states $0^{+-}$, $2^{+-}$, and $3^{--}$. It is worth mentioning the results predicted by the single pole (SP) and dipole (DP) Regge model \cite{Szanyi:2019kkn}, which are fitted by high-energy $pp$ scattering: for the SP Regge model, the predicted value of the $2^{++}$ glueball mass is 1.747 GeV; for the DP Regge model, the masses of the $2^{++}$ glueball and the $3^{--}$ oddball are 1.758 GeV and 3.001 GeV, respectively. These predictions are lower than the results of the holographic model, but still in an acceptable range. This may indicate that the $2^{++}$ glueball and the $3^{--}$ oddball are mixed with quark states, both of which are hybrid glueball/oddball states.

\subsection{DhQCD for light flavor hadron spectra}
In this section we will try to derive a self-consistent holographic model, which could describe the linear Regge behavior in the light meson spectra and the linear quark potential. We will start from the general system Eq.\eqref{action-full}. Since we will focus on the vacuum, we will neglect the $U(1)$ gauge field $\mathcal{F}_{mn}$. Also, in the vacuum, only the metric, the dilaton filed, and the expectation value of the scalar field $\chi$ is non-vanishing. Thus, for the background field, we could get the following effective action
\begin{equation}
S_{vac} = S_{G,vac}+ \frac{N_f}{N_c} S_{S,vac},
\end{equation}
with
\begin{eqnarray}
&&\!\!\!\!\!\!\!\!\!\!\!\!\!\!\!\! S_{G,vac} = \frac{1}{16\pi G_5}\int d^5x\sqrt{g_s}e^{-2\Phi}\big(R
\nonumber\\
&&\qquad\qquad\qquad\qquad+4\partial_M\Phi\partial^M\Phi-V_G(\Phi)\big), \\
 &&\!\!\!\!\!\!\!\!\!\!\!\!\!\!\!\! S_{S,vac} = - \int d^5x \sqrt{g_s}e^{-\Phi}(\frac{1}{2}\partial_M\chi
 \partial^M \chi+V_{C}(\chi,\Phi)).
\end{eqnarray}
Here, $V_C(\chi,\Phi)$ is the potential of the scalar field, and we have taken $\lambda=\frac{N_f}{N_c}$ with $N_f=2$, $N_c=3$ in this section. By the redefinition of the dimensional fields as $ L^{\frac{3}{2}}\chi\rightarrow
\chi,L^3V_C\rightarrow V_C, \frac{16\pi
G_5 N_f}{L^3 N_c}\rightarrow \lambda $, the vacuum action becomes
\begin{eqnarray}
S_{vac} \!\!\! &=& \!\!\! \frac{1}{16 \pi G_5} \int d^5 x \sqrt{g_s} \big\{ e^{-2\Phi}[R
+4\partial_M\Phi \partial^M \Phi - V_G(\Phi)] \nonumber\\
&-& \lambda e^{-\Phi} (\frac{1}{2} \partial_M\chi \partial ^M \chi
+ V_C(\chi,\Phi))\big\}.
\end{eqnarray}
At zero temperature, one could take the following metric ansatz
\begin{eqnarray}\label{metric-ansatz}
g^s_{MN}=b_s^2(z)(dz^2+\eta_{\mu\nu}dx^\mu dx^\nu), ~ ~ b_s(z)\equiv e^{A_s(z)}.
\end{eqnarray}
Then the equation of motion could be derived as
\begin{eqnarray}
 -A_s^{''}+A_s^{'2}+\frac{2}{3}\Phi^{''}-\frac{4}{3}A_s^{'}\Phi^{'}
 -\frac{\lambda}{6}e^{\Phi}\chi^{'2}&=&0, \label{Eq-As-Phi} \\
 \Phi^{''}+(3A_s^{'}-2\Phi^{'})\Phi^{'}-\frac{3\lambda}{16}e^{\Phi}\chi^{'2}&&\nonumber\\
 -\frac{3}{8}e^{2A_s-\frac{4}{3}\Phi}\partial_{\Phi}\left(V_G(\Phi)
 +\lambda e^{\frac{7}{3}\Phi}V_C(\chi,\Phi)\right)&=&0, \label{Eq-VG}\\
 \chi^{''}+(3A_s^{'}-\Phi^{'})\chi^{'}-e^{2A_s}V_{C,\chi}(\chi,\Phi)&=&0. \label{Eq-Vc}
\end{eqnarray}

\begin{figure}[h]
\begin{center}
\includegraphics[width=0.4\textwidth]{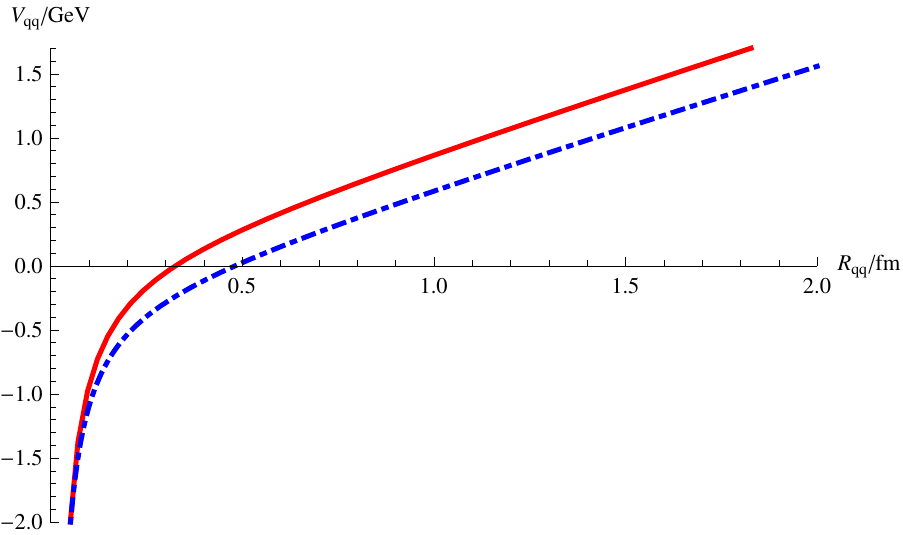} \hspace*{0.1cm}
\hskip 0.15 cm
\textbf{( a ) }\\
\includegraphics[width=0.4\textwidth]{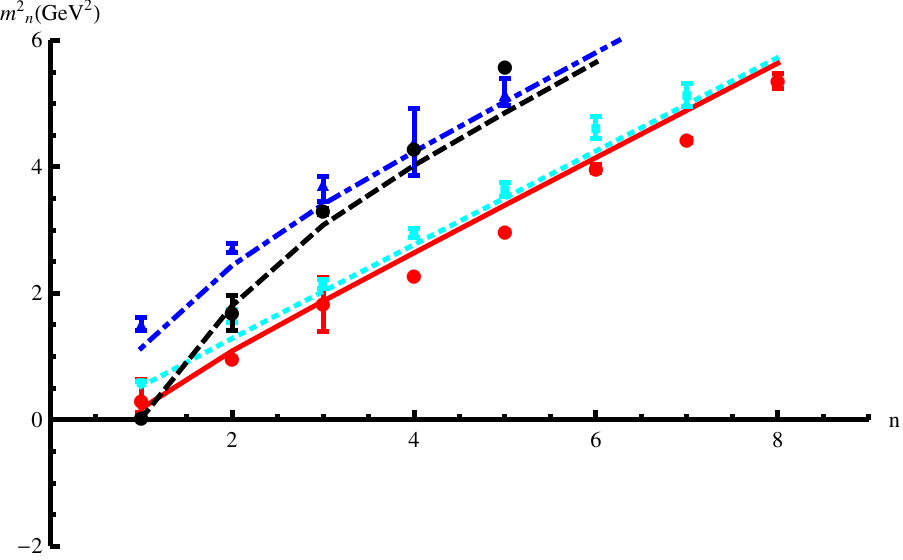}
 \hskip 0.15 cm 
 \textbf{( b )} \\
\end{center}
\caption[]{
Results of the quark-antiquark potential and the light meson spectra from the DhQCD model, where we have taken the model IA in Ref.~\cite{Li:2013oda}. (a)The quark-antiquark potential in the DhQCD model(red solid line) compared with the Cornell potential(blue dot-dashed line). (b)The light meson spectra in the DhQCD model. The figures are from Ref.~\cite{Li:2013oda}. }
 \label{potential-spectra}
\end{figure}

Phenomenologically, to get a consistent description of the light flavor physics, there are several constraints to be satisfied. Firstly, the IR behavior of the dilaton filed is related to the Regge slope. Secondly, to produce a linear quark potential, the derivative of the warp factor $A_s$ should equal zero at certain $z_c$ or approaching zero when $z$ goes to infinity. Thirdly, $\frac{e^{2A_s}}{z^2}\chi^2$ is responsible for the mass splits of chiral partners, thus it should not be zero at IR. Considering all these requirements, we take 
\begin{eqnarray}
\Phi(z)&&=\mu_G^2z^2 \label{phiz}\\
\chi^{'}(z)&&=\sqrt{8/\lambda}\mu_G e^{-\Phi/2}(1+c_1 e^{- \Phi}+c_2
e^{-2\Phi}),\label{chiz}
\end{eqnarray}
with $c_1=-2+\frac{5\sqrt{2\lambda}m_q\zeta}{8\mu_G}+\frac{3\sqrt{2\lambda}\sigma}{4\zeta
\mu_G^3},c_2=1-\frac{3\sqrt{2\lambda}m_q\zeta}{8\mu_G}-\frac{3\sqrt{2\lambda}\sigma}{4\zeta\mu_G^3}$. After taking $m_q=5.8 {\rm MeV},\sigma=(180 {\rm MeV})^3, \mu_G=0.43\rm{GeV}, G_5/L^3=0.75$ (the model IA in Ref.~\cite{Li:2013oda}), we solve the background field $A_s,\Phi,f,\chi$ from the equation of motion. Then we obtain the quark-antiquark potential by solving the Wilson loop from holographic dictionary. The results are given in Fig.\ref{potential-spectra}(a). From the figure we can see that in such a model the quark-antiquark potential is linear at large distance. Qualitatively and even quantitatively it agrees with the Cornell potential. 

Then, we can consider the mesonic excitation on the vacuum, which is dual to the corresponding perturbative modes. For the scalar channel, we have $X=(\frac{\chi}{2}+s)e^{i 2\pi^a t^a}$, with $s, \pi$ the scalar and pseudo-scalar perturbation respectively. For the vector channel, we have the vector perturbation
$v_\mu$ and axial vector perturbation $a_\mu$. From the full action, one can easily derive the equation of motion for $s,\pi,v_\mu,a_\mu$ as
\begin{eqnarray}\label{scalar-sn}
& & -s_n^{''}+V_s(z)s_n=m_n^2s_n,  \\
& & -\pi_n''+V_{\pi,\varphi}\pi_n=m_n^2(\pi_n-e^{A_s}\chi\varphi_n),  \\
& &  -\varphi_n''+ V_{\varphi} \varphi_n=g_5^2 e^{A_s}\chi(\pi_n-e^{A_s}\chi\varphi_n), \\
& & -v_n^{''}+V_v(z)v_n=m_{n,v}^2v_n, \label{vector-n} \\
& & -a_n^{''}+V_a a_n = m_n^2 a_n,
\end{eqnarray}
with the schrodinger-like potentials
\begin{eqnarray}
&&V_s=\frac{3A_s^{''}-\phi^{''}}{2}+\frac{(3A_s^{'}-\phi^{'})^2}{4}+e^{2A_s}V_{C,\chi\chi},
\label{s-vz}\\
&& V_{\pi,\varphi}=\frac{3A_s^{''}-\phi^{''}+2\chi^{''}/\chi-2\chi^{'2}/\chi^2}{2} \label{ps-vz} \nonumber \\
 & & ~~~~~~~~~~ +\frac{(3A_s^{'}-\phi^{'}+2\chi^{'}/\chi)^2}{4},  \\
&& V_{\varphi} = \frac{A_s^{''}-\phi^{''}}{2}+\frac{(A_s^{'}-\phi^{'})^2}{4},  \\
&& V_v=\frac{A_s^{''}-\phi^{''}}{2}+\frac{(A_s^{'}-\phi^{'})^2}{4} ~, \label{v-vz} \\
&& V_a = \frac{A_s^{'}-\phi^{'}}{2}+\frac{(A_s^{'}-\phi^{'})^2}{4}+g_5^2 e^{2A_s}\chi^{2}
\label{a-vz}.
\end{eqnarray}
Then, the eigenvalues of those Schrodinger-like equations could be identified as the corresponding mesons' masses. The results are shown in Fig.\ref{potential-spectra}(b). There we could see that with only four parameters the DhQCD model could match the experimental data of light mesons quite well. The linear behavior in the higher excitations is reproduced. The mass splits of the chiral partners are well described. Furthermore, by self-consistently solving the system, the linear potential and the linear spectra are realized simultaneously.

\subsection{DhQCD for heavy flavor hadron spectra}
In this subsection, the $N_f=4$ holographic QCD model describing both light and heavy mesons is introduced, which is similar to the holographic soft wall model \cite{Karch:2006pv} and includes the complex scalar field $X$, two gauge fields $L_M^a$ and $R_M^a$.
The difference is that the action contains not only the dilaton field, which can describe the linear Regge behavior, but also the hard cutoff $z_m$, which enables the splitting of the Regge slopes of light and heavy mesons. 
Furthermore, an additional scalar field $H$ need to be introduced to distinguish between vector $\rho$ and $J/\psi$ mesons, which are analogous to the $\Psi$ fields \cite{Liu:2016iqo,Liu:2016urz} of the D4-D8 model, to characterize the large mass gap between light and heavy quarks.

Due to the difficulty of solving for the total action Eq.\eqref{action-full}, the calculations in this subsection are under the probe approximation and set the background metric to AdS$_5$. In summary, the total action describing the light and heavy mesons is	\begin{eqnarray}\label{action-heavy}
		S_M &=& -\int_\epsilon^{z_m} d^5 x \sqrt{-g} ~ e^{-\phi} ~{\rm Tr} \Big \{ (D^M X)^{\dag} (D_M X) \nonumber\\
		&&+ m_5^2 |X|^2 +(D^M H)^{\dag} (D_M H)  + m_5^2 |H|^2 \nonumber\\
		&&+ \frac{1}{4 g_5^2} \left ( L^{MN} L_{MN} + R^{MN} R_{MN} \right )
		\Big \} \, ,
	\end{eqnarray}
	where the covariance derivatives of the $H$-field and $X$-field are defined as $D_{M} X=\partial_M X -iL_M X +i X R_M$ and $D_M H = \partial_M H - i V_M^{15} H - i H V_M^{15}$, respectively.
	According to the AdS/CFT dictionary, the five-dimensional mass square of both scalar fields is set to $m_5^2=\Delta(\Delta-4)=-3$.
	The strength tensors of the left gauge field $L_{M}$ and right gauge field $R_{M}$ are written as
	\begin{eqnarray}
		L_{MN} &=& \partial_M L_N - \partial_N L_M - i \left [ L_M , L_N \right ], \nonumber\\
		R_{MN} &=& \partial_M R_N - \partial_N R_M - i \left [ R_M , R_N \right ],
	\end{eqnarray}
	where $L_M=L_M^at^a$ and $R_M=R_M^at^a$ with the generators $t^a$ of the $SU(4)$. With reference \cite{Erlich:2005qh}, the five-dimensional coupling constant $g_5^2$ is fixed to $g_5^2=12\pi^2/N_c$ by current algebra.
	The scalar field $X$ is decomposed according to the following form
	\begin{eqnarray}
		X = e^{i\pi^at^a} \, X_0 \, e^{i\pi^bt^b},
	\end{eqnarray}
	with $X_0={\rm diag}[v_l(z),v_l(z),v_s(z),v_c(z)]$. 
	Based on the AdS/CFT correspondence, the expansion of the field at the conformal boundary yields the source and expectation values of its dual operator, and therefore, $v_{l,s,c}(z)$ in $X_0$ has asymptotic behavior $v_{l,s,c}(z)\to M_{l,s,c}z+\Sigma_{l,s,c}z^3$ at UV boundary. 
	Since only the effects of heavy quarks are described, another scalar field $H$ has $H={\rm diag}[0,0,0,h_c(z)]$. 
	Similarly, at the UV boundary, the expansion of $h_c$ takes the form $h_c(z)\to m_c z$. 
	It should be noted that $m_c\neq M_c$ was considered in order to obtain a better mesons spectra. 
	
	By expanding the action Eq.(\ref{action-heavy}) and according to the AdS/CFT duality, the meson mass and decay constants, which come from the quadratic term of the action, and the three- and four-point coupling constants, which come from the cubic and quartic terms of the action, respectively, can be obtained. 
	
	The equation of motion for the scalar field can be written as
	\begin{eqnarray}\label{eom-v}
		-\frac{z^3}{e^{-\phi(z)}}\partial_z \frac{e^{-\phi}}{z^3}\partial_z v_q(z)+\frac{m_5^2}{z^2}v_q(z)=0.
	\end{eqnarray}
	The analytic solutions are
	\begin{eqnarray}
		v_q(z)&=&C_1(q)~z~\sqrt{\pi}~U(\frac{1}{2},0,\phi)\nonumber\\
		&&-C_2(q)~z~L(-\frac{1}{2},-1,\phi),
	\end{eqnarray}
	with the confluent hypergeometric function $U$, the generalized Laguerre polynomial $L$ and two constants $C_1(q)$ and $C_2(q)$.
	The equations of motion for the pseudoscalar, vector, and axial vector meson wave functions are shown below
	\begin{eqnarray}
		&&\left(-\frac{z}{e^{-\phi}}\partial_z\frac{e^{-\phi}}{z}\partial_z+\frac{2g_5^2(m_V^{ab}-M_V^{ab})}{z^2}\right) V^a_{\mu\perp}(q,z)\nonumber\\
		&&=-q^2V^a_{\mu\perp}(q,z), \label{eomv}\\
		&&\left(-\frac{z}{e^{-\phi}}\partial_z\frac{e^{-\phi}}{z}\partial_z+\frac{2g_5^2M_A^{ab}}{z^2}\right) A^a_{\mu\perp}(q,z)\nonumber\\
		&&=-q^2A^a_{\mu\perp}(q,z), \label{eoma}\\
		&&q^2\partial_z \varphi^a(q,z)+\frac{2g_5^2M_A^{ab}}{z^2}\partial_z \pi^a(q,z)=0\,,\label{eom-pi-1}\\			
		&&\frac{z}{e^{-\phi}}\partial_z \left(\frac{e^{-\phi}}{z}\partial_z\varphi^a(q,z) \right)\nonumber\\
		&&-\frac{2g_5^2M_A^{ab}}{z^2}\left(\varphi^a(q,z)-\pi^a(q,z)\right)=0\,,\label{eom-pi-2} 					
	\end{eqnarray}
	where $V$, $A$ and $\varphi$, $\pi$ denote the vector, axial vector and pseudoscalar, respectively. 
	Through the two-point function, the decay constants of the mesons are
	\begin{eqnarray}
		F_{V^n}^2&=&\frac{[e^{-\phi(\epsilon)}\psi_{V^n}^\prime(\epsilon)/\epsilon]^2}{g_5^2}|_{\epsilon\to 0},\\
		F_{A^n}^2&=&\frac{[e^{-\phi(\epsilon)}\psi_{A^n}^\prime(\epsilon)/\epsilon]^2}{g_5^2}|_{\epsilon\to 0},\\
		f_{\pi}^2&=&-\frac{e^{-\phi(\epsilon)}\partial_zA(0,\epsilon)}{\epsilon~ g_5^2}|_{\epsilon\to 0}.
	\end{eqnarray}
	The coupling constants between the mesons can be obtained through the three- and four-point functions as
    \begin{eqnarray}\label{3cc-1}
		g_{VVV}&=&\int_0^{z_m} dz\frac{e^{-\phi(z)}}{2g_5^2z}f^{bca}\psi_{V^{(n)}}^a\psi_{V^{(m)}}^b\psi_{V^{(k)}}^c,\\
		g_{VAA}&=&\int_0^{z_m}dz\frac{e^{-\phi(z)}}{2g_5^2z}f^{bca}\psi_{V^{(n)}}^a\psi_{A^{(m)}}^b\psi_{A^{(k)}}^c,\label{3cc-2}\\
		g_{VA\pi}&=&\int_0^{z_m} dz\frac{e^{-\phi(z)}}{z^3}2\psi_{V^{(m)}}^a\psi_{A^{(m)}}^b\nonumber\\
		&&\psi_{\pi^{(k)}}^c(g^{bac}-h^{abc}),\label{3cc-3}\\
		g_{V\pi\pi}&=&\int_0^{z_m} dz\frac{e^{-\phi(z)}}{z^3}\psi_{V^{(n)}}^{a}\psi_{\pi^{(m)}}^b\nonumber\\
		&&\psi_{\pi^{(k)}}^c(h^{abc}+h^{acb}-2g^{cab}),\label{3cc-4}
	\end{eqnarray}
	and
	\begin{eqnarray}
    	g_{VVVV}&=&\int_0^{z_m} dz\frac{e^{-\phi(z)}}{4g_5^2z}f^{abcd}\psi_{V^{(n)}}^a\psi_{V^{(m)}}^b\nonumber\\
    	&&\psi_{V^{(k)}}^c\psi_{V^{(j)}}^d,\label{4cc-1}\\
    	g_{VVAA}&=&\int_0^{z_m} dz\frac{e^{-\phi(z)}}{4g_5^2z}2\psi_{V^{(n)}}^a\psi_{V^{(m)}}^b\psi_{A^{(k)}}^c\nonumber\\&&\psi_{A^{(j)}}^d(f^{abcd}+f^{cbad}),\label{4cc-2}\\
    	g_{AAAA}&=&\int_0^{z_m} dz\frac{e^{-\phi(z)}}{4g_5^2z}f^{abcd}\psi_{A^{(n)}}^a\psi_{A^{(m)}}^b\nonumber\\
    	&&\psi_{A^{(k)}}^c\psi_{A^{(j)}}^d,\label{4cc-3}\\
    	g_{VV\pi\pi}&=&\int_0^{z_m} dz\frac{e^{-\phi(z)}}{z^3}\psi_{V^{(n)}}^a\psi_{V^{(m)}}^b\nonumber\\
    	&&\psi_{\pi^{(k)}}^c\psi_{\pi^{(j)}}^d(h^{abcd}-g^{acbd}),\label{4cc-4}\\
    	g_{AA\pi\pi}&=&\int_0^{z_m} dz\frac{e^{-\phi(z)}}{z^3}\psi_{A^{(n)}}^a\psi_{A^{(m)}}^b\nonumber\\
    	&&\psi_{\pi^{(k)}}^c\psi_{\pi^{(j)}}^d(k^{acbd}-l^{abcd}),\label{4cc-5}\\
    	g_{A\pi\pi\pi}&=&\int_0^{z_m} dz\frac{e^{-\phi(z)}}{z^3}\psi_{A^{(n)}}^a\psi_{\pi^{(m)}}^b\psi_{\pi^{(k)}}^c\nonumber\\
    	&&\psi_{\pi^{(j)}}^d(l^{bacd}+\frac{l^{abcd}}{3}+\frac{l^{acbd}}{3}\nonumber\\
    	&&+\frac{l^{acdb}}{3}-k^{bcad}-k^{cbad}),\label{4cc-6}\\
    	g_{\pi\pi\pi\pi}&=&\int_0^{z_m} dz\frac{e^{-\phi(z)}}{z^3}(\psi_{\pi^{(n)}}^a\psi_{\pi^{(m)}}^b\psi_{\pi^{(k)}}^c\psi_{\pi^{(j)}}^d\nonumber\\
    	&&+\psi_{\pi^{(n)}}^{\prime,a}\psi_{\pi^{(m)}}^{\prime,b}\psi_{\pi^{(k)}}^c\psi_{\pi^{(j)}}^d)\nonumber\\
    	&&(\frac{k^{acbd}+k^{cabd}+k^{acdb}+k^{cadb}}{4}\nonumber\\
    	&&-\frac{l^{abcd}+l^{acbd}+l^{acdb}}{3}),\label{4cc-7}
    \end{eqnarray}
    where $\psi$ is the eigenwave function given by the equation of motion. 
    Here, $M_A^{ab}$, $M_V^{ab}$, $m_V^{ab}$, $h^{abc}$, $g^{abc}$, $g^{abcd}$, $l^{abcd}$, $h^{abcd}$, $k^{abcd}$, and $f^{abcd}$ appearing in the equations of motion and coupling constants are defined as
    \begin{eqnarray}
		&M_A^{ab}={\rm Tr}(\{t^a,X_0\}\{t^b,X_0\}),\\ 
		&M_V^{ab}={\rm Tr}([t^a,X_0][t^b,X_0]),\\
		&m_V^{15,15}={\rm Tr}(\{H,t^{15}\}\{H,t^{15}\}),\\
		&f^{abcd}=f^{\alpha ab}f^{\alpha cd},\\
		&h^{abc}=i~{\rm Tr} \left([t^a,X_0]\{t^b,\{t^c,X_0\} \}\right),\\ &g^{abc}=i~{\rm Tr}(\{t^a,X_0\}[t^b,\{t^c,X_0\}]),\\
		&h^{abcd}={\rm Tr}([t^a,X_0][t^b,\{t^c,\{t^d,X_0\}\}]),\\
		&g^{abcd}={\rm Tr}([t^a,\{t^b,X_0\}][t^c,\{t^d,X_0\}]),\\
		&l^{abcd}={\rm Tr}\left(\{t^a,X_0\}\{t^b,\{t^c,\{t^d,X_0\}\}\} \right),\\
		&k^{abcd}={\rm Tr}(\{t^a,\{t^b,X_0\}\} \{t^c,\{t^d,X_0\}\}),
	\end{eqnarray}
	with structure constant $f^{abc}$ of $SU(4)$ Lie algebra. 
	
\begin{figure}[!thb]
    	\centering
    	\includegraphics[width=0.45\textwidth]{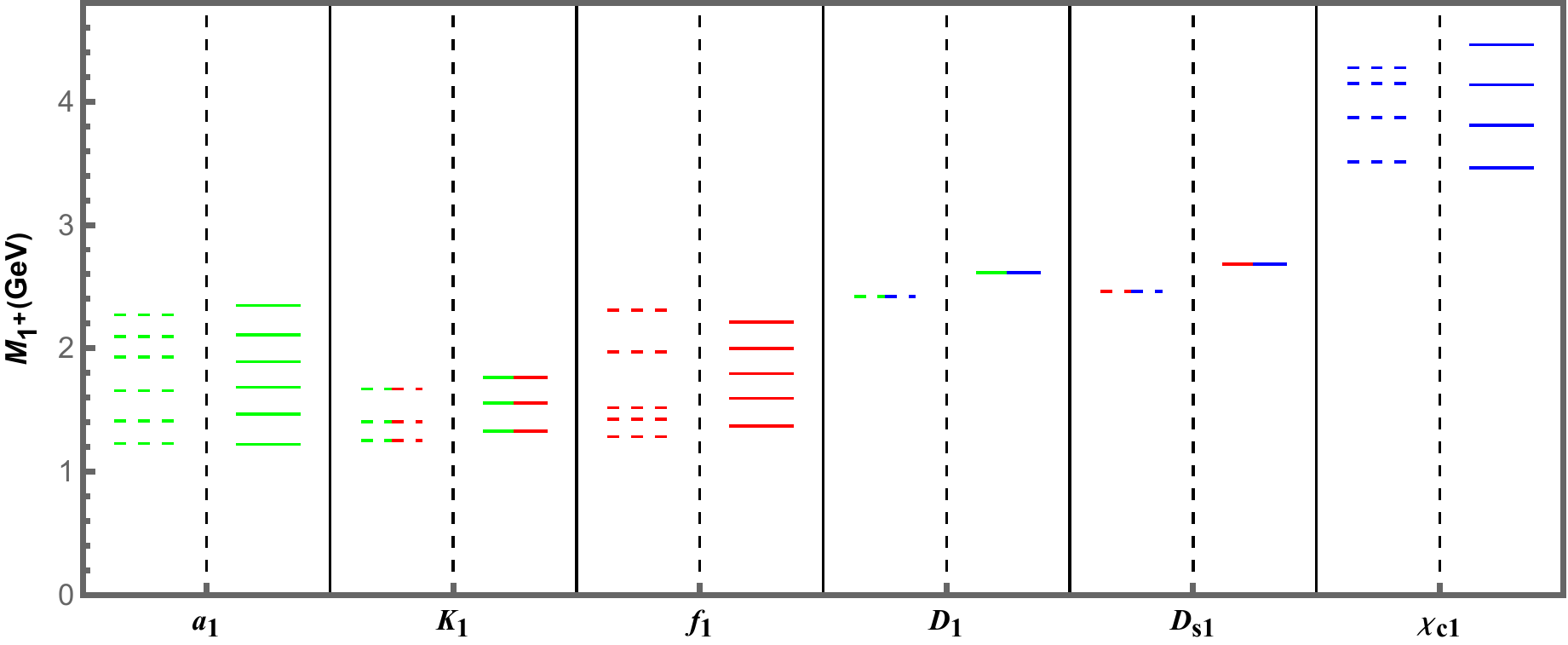}
    	\includegraphics[width=0.45\textwidth]{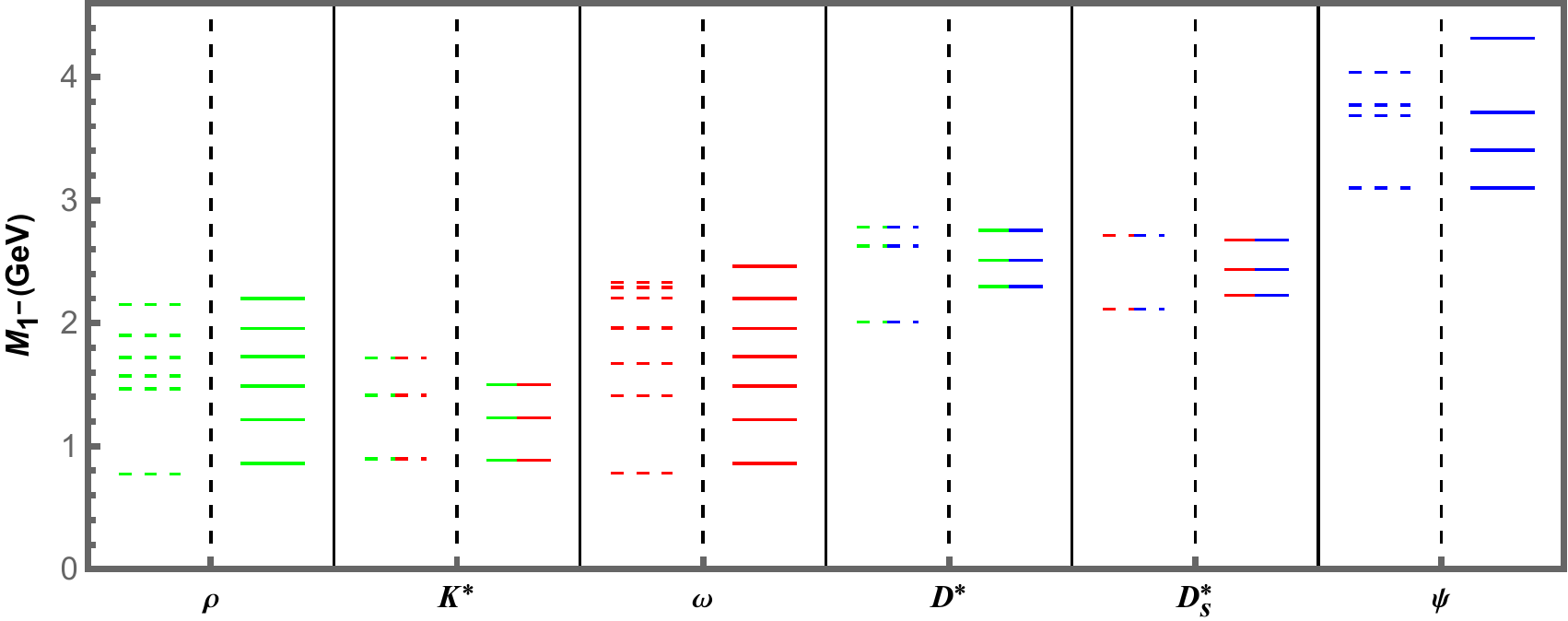}
    	\includegraphics[width=0.45\textwidth]{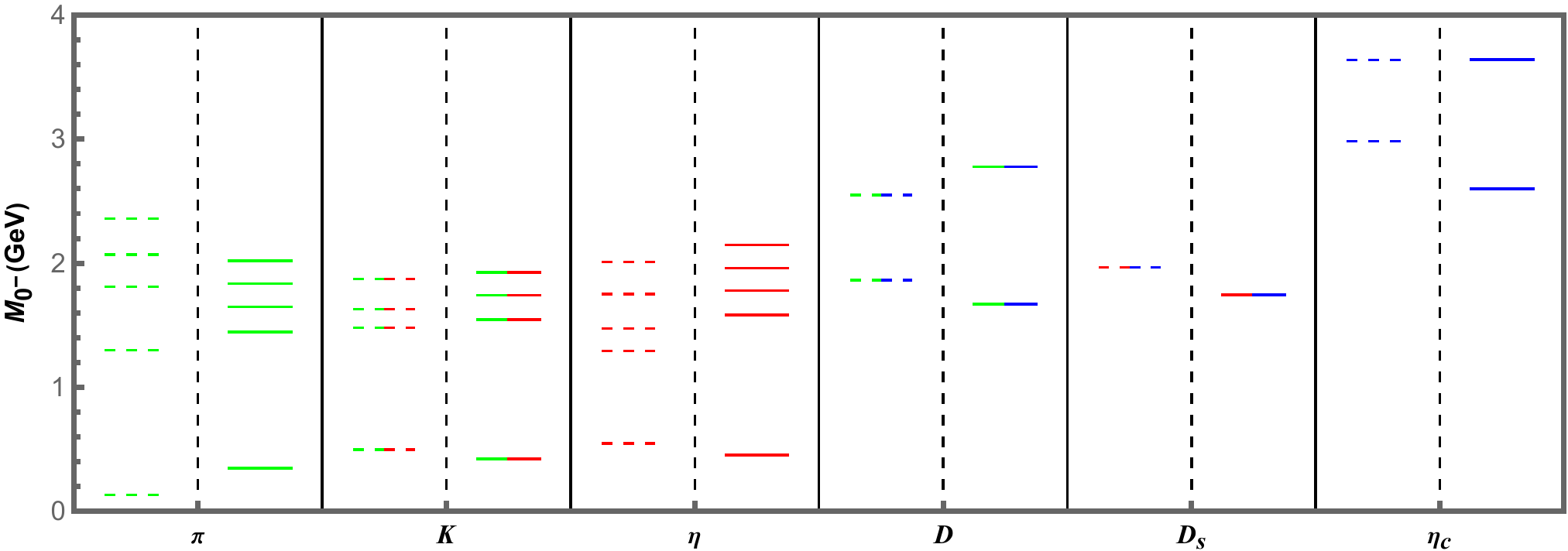}
    	\caption{The mesons spectra of the holographic QCD model with $N_f=2+1+1$, where the dashed line represents the experimental data, selected from the PDG \cite{ParticleDataGroup:2020ssz}, while the solid line indicates the results of the model. The different colors are used in the figure to denote the different constituent quarks in the mesons, with red, green and blue representing $ud$, $s$ and $c$ quarks, respectively. The figures are from Ref. \cite{Chen:2021wzj}.}
    	\label{fig:mesonspectra}	
    \end{figure}

	The final mesons spectra of the model are shown in Fig.\ref{fig:mesonspectra}, where the dashed lines denote the experimental data, chosen from the PDG \cite{ParticleDataGroup:2020ssz}, and the solid lines indicate the numerical results. The red, green and blue in the figure represent $u,d$, $s$ and $c$ quarks, respectively, which are used to identify the different constituent quarks of the mesons.
	
    In the holographic QCD model describing the heavy flavors, there are six physical parameters, including the quadratic term coefficients $\mu$ of the dilaton field and the IR cutoff $z_m$, which are used to fit the Regge slope of the light and heavy mesons, the vacuum expectation value $C_1(q)$ of the four flavor quarks, i.e., the $u,d$, $s$, and $c$ quarks, with $q=(l,s,c)$, and the constant $D_1$ of the auxiliary field $H$, which characterizes the mass difference between the light and heavy quarks.
    
    The above six parameters are fitted by the following process. the quadratic term coefficient $\mu$ of the dilaton field is chosen to be $0.43$GeV, considering that the ground state mass of the vector meson $\rho$ depends only significantly on it, at which point the mass of the $\rho$ meson and the Regge slope can be fitted simultaneously. 
    After fixing the value of $\mu$, the vacuum expectation value $C_1(q)$ of the four flavor quarks is fitted using the masses and the Regge slope of the meson $a_{1}$, $K^{_{*}}$ and $\chi_{c1}$, respectively. 
    The last two parameters, $z_m$ and $D_1$, which are associated with heavy mesons, are chosen based on the mass of the $J/\psi$ meson and its Regge slope. For better numerical results, these two parameters are finally determined by the masses of the $J/\psi$ and $\psi(3770)$ mesons.
    Applying the method of expanding the field $X$ at the conformal boundary, the parameters $C_1(q)$ can be transformed into quark masses and condensations, and the final results are $M_{l}=140~$MeV, $M_{s}=200~$MeV, $M_{c}=1200~$MeV and $\Sigma_{l}=(135~\rm{MeV})^{3}$, $\Sigma_{s}=(152~\rm{MeV})^{3}$, $\Sigma_{c}=(276~\rm{MeV})^{3}$.
    Similarly, the parameter $D_1$ indicates $m_c=1020~$MeV and $\sigma_{c}=(262~\rm{MeV})^{3}$. It should be noted that the QCD phenomenology suggests $\langle\bar{s}s\rangle\sim 0.8 \langle\bar{l}l\rangle$ and $\langle\bar{c}c\rangle\sim 0$, which is not consistent with the results of the holographic model. The problem may come from the soft wall model itself, since the two integration constants $C_1$ and $C_2$ of the equations of motion of the scalar field are not chosen arbitrarily. As described in \cite{Li:2013oda}, since $C_2$ leads to nonlinear behavior of the $a_1$ meson spectra that must be set to $0$, the only integration constant $C_1$ determines the quark mass and condensation, making the value of condensation proportional to the quark mass. The solution is to consider the backreaction of the scalar field and to solve the probe and the background together, i.e., to solve the Einstein-dilaton-scalar system, as in \cite{Li:2013oda}.
    
    In the above way, the parameters of the model are fixed and the masses of the pseudoscalar, vector and axial vector mesons and their excited states can be obtained, as shown in the solid line in Fig.\ref{fig:mesonspectra}.
    The calculations of the holographic model show that the results for the axial vector $a_1$ and $K_1$ mesons are in good agreement with the data; the results for $D_1$ and $D_{s1}$ mesons are about $0.2$ GeV heavier than the experimental data; the ground state and the first two excited states of $\chi_{c1}$ mesons are good, while their third excited state is about $0.2$ GeV heavier than the data.
    It is worth noting that the results for $f_1$ mesons deviate slightly from the data because the mixing of $s$ and $u, d$ quarks is not considered.
    
    For vector mesons, the excited state results for $K^{*}$ mesons are about $0.2$ GeV lighter than the experimental data, which can be obtained from the deviation of their Regge behavior, probably because the masses of $s$ quarks and $u, d$ quarks are too close in the model; the ground state masses of $D^{*}$ mesons are about $0.3$ GeV heavier than the data, while the excited states are relatively better; the first and third excited states of the $\psi$ mesons do not match the data very well. 
    For light vector mesons, the results are degenerate, i.e., no distinction can be made between $\rho$ and $\omega$ mesons. In principle, the distinction can be achieved by adding additional parameters to the auxiliary $H$-field. As can be seen from the specific values, the results are closer to $\rho$ mesons than to $\omega$ mesons, due to the omission of the mixing of $s$ quarks with $u, d$ quarks. As for the selection of $\omega$ mesons instead of $\phi$ mesons, the reason is that the equations of motion of vector mesons are closer to the pure $u\bar{u},d\bar{d}$ states. 
    
    It is notable that the mass of the experimental $D_{s1}^*$ (2700) state is close to that of the second excited state of the $D_{s1}$ in the holographic model. 
    Considering that the Regge slope depends only on the constituent quarks of the mesons, it is conjectured that a new excited state may exist between $D_{s1}^*$(2700) and $D_{s1}$, with a mass of about 2436 MeV predicted by the model.
    
    From the results, it can be seen that the excited state mass of the pseudoscalar $\eta$ meson is about $0.25$ GeV heavier than the data, but its Regge behavior is still maintained, probably due to the missing mixing term as in the case of the $\omega$ meson; the ground state masses of the $D$ and $D_s$ mesons are $0.2$ GeV lighter than the data; the ground state mass of $\eta_c$ mesons is $0.4$ GeV lighter compared to the data. 
    It is known that $\eta^\prime$ mesons are associated with chiral anomalies, so $\eta$ mesons rather than $\eta^\prime$ mesons are considered in the model. For the $\pi$ meson, its second, third and fourth excited states are approximately $0.15$-$0.3$ GeV lighter than the data, and thus its Regge slope does not match. 
    Note that the Regge slope of the $\pi$ meson in the experimental data is not consistent with that of the other light mesons, so there are still difficulties in realizing this difference in the holographic model.
    
    The holographic model not only introduces the dilaton field, but the additional IR cutoff $z_m$ and the auxiliary field $H$ are also included, where the dilaton serves to realize the linear Regge trajectory, the IR cutoff is used to distinguish the Regge slope of the light and heavy mesons, and the $H$ field to improve the intercept of the linear behavior of the heavy mesons. 
    In addition, the decay constants of mesons, three-point and four-point couplings are also calculated in the model, and the specific results can be referred to \cite{Chen:2021wzj}.

\section{QCD phase transitions, thermodynamical and transport properties}
\subsection{Decnfinement phase transition, thermodynamical and transport properties}
\label{sec-eos-transport}
In this section, we will focus on the deconfinement phase transition and the thermodynamic quantities at finite temperatures from the dynamical holographic QCD model. Since we would not consider baryonic density here, the $U(1)$ gauge filed in the full DhQCD model could be neglected. Furthermore, since the centre symmetry is an exact symmetry only in the gluon system, in which the deconfinement phase transition could be well described, we will focus on the contributions from the glue-sector in this subsection. Thus, the full system would be reduced to  Eq.\eqref{action-ED-Einstein} in the Einstein frame or Eq.\eqref{action-ED-string} in the string frame. We will consider this scenario as a quenched DhQCD model.

From the holographic dictionary, an usual way to introduce temperature is to consider the black hole solutions and identify the Hawking temperature as the temperature of the 4D field system. Therefore, we take the metric ansatz at finite temperature in the string frame as the following form
\begin{equation} \label{metric-stringframe}
ds_S^2=
e^{2A_s}\left(-f(z)dt^2+\frac{dz^2}{f(z)}+dx^{i}dx^{i}\right).
\end{equation}
The warp factor $A_s$, the blackening factor $f(z)$, and the dilaton field $\Phi(z)$ satisfies the following equation
\begin{center}
\begin{eqnarray}
 &&-A_s^{''}+A_s^{'2}+\frac{2}{3}\Phi^{''}-\frac{4}{3}A_s^{'}\Phi^{'}=0, \label{Eq-As-Phi-T} \\
 &&f''(z)+\left(3 A_s'(z) -2 \Phi '(z)\right)f'(z)=0,\label{Eq-As-f-T}\\
 &&\frac{8}{3} \partial_z
\left(e^{3A_s(z)-2\Phi} f(z)
\partial_z \Phi\right)-
e^{5A_s(z)-\frac{10}{3}\Phi}\partial_\Phi V_E=0,\nonumber\\
\end{eqnarray}
\end{center} 
which are derived from the Einstein equations and the equation of motion for the dilaton field.
Once those equations are solved out, the temperature of the black hole solution could be obtained as 
\begin{equation} \label{temp-general}
T =-\frac{f^\prime(z_h)}{4\pi},
\end{equation}
with $z_h$ the horizon where $f(z_h)=0$. From Eq.\eqref{Eq-As-f-T}, it is easy to get
\begin{eqnarray} \label{solu-f}
f(z)= 1- f_{c}^h \int_0^{z} e^{-3A_s(z^{\prime})+2\Phi(z^{\prime})} dz^{\prime},
\end{eqnarray}
with
\begin{eqnarray}\label{fc}
f_{c}^h= \frac{1}{\int_0^{z_h} e^{-3A_s(z^{\prime})+2\Phi(z^{\prime})} dz^{\prime} }.
\end{eqnarray}
Therefore, one has
\begin{equation} \label{temp-sol}
T =\frac{e^{-3A_s(z_h)+2\Phi(z_h)}}{4\pi \int_0^{z_h} e^{-3A_s(z^{\prime})+2\Phi(z^{\prime})} dz^{\prime} }.
\end{equation} 
Meanwhile, the entropy could be identified with the Beikenstein-Hawking entropy, and its density is related to the area $A$ of the black hole horizon through the following equation
\begin{equation}\label{entropy-area}
s=\frac{A}{4G_5 V_3}=\frac{L^3}{4G_5 }\left(\frac{e^{A_s-\frac{2}{3}\phi}}{z}\right)^3_{z=z_h},
\end{equation}
Here, $V_3$ is the volume of the system. With the expressions of the entropy density and temperature, one could derive all other thermodynamic quantities through the general thermodynamic relations. For example, one could get the pressure $p$, energy denstiy $\epsilon$ through the following relations
\begin{eqnarray}
\frac{dp(T)}{dT}&=&s(T),\label{pressure}\\
\epsilon&=&-p+sT.\label{energy}
\end{eqnarray}

For the soft-wall model, the quadratic dilaton field is responsible for the linear Regge trajectories. So we assume that for the pure gluon system the dilaton profile should take a similar form, i.e. $\Phi=\mu_G z^2$, with $\mu_G$ a mass scale. Then one can solve the equations and get an analytic solution 
of the metric prefactor
\begin{equation}\label{As-sol}
A_s(z) =\log(\frac{L}{z})-\log(_0F_1(5/4,\frac{\mu_G^4z^4}{9}))+\frac{2}{3}\mu_G^2z^2,
\end{equation}
with $L$ the AdS radius. Then one can obtain the results of temperature, entropy density, and the other thermodynamical quantities. It is found that the black hole solutions appear only above a certain temperature $T_c$, indicating the appearance of a phase transition. As shown in Ref.\cite{Li:2011hp}, the Polyakov loop, which is the order parameter of the deconfinement phase transition, becomes 
discontinuous near $T_c$. Thus, the geometric phase transition should be considered as the deconfinement phase transition, just as expected from the model building. By fitting the critical temperature $T_c$ for the deconfinement phase transition from the quenched lattice results, we take $\mu_G=0.75 {\rm GeV}$ and get $T_c=255 {\rm MeV}$. Then we can extract all other thermodynamic quantities numerically and compare the results with lattice data.

\begin{figure}
  \begin{center}
    \includegraphics[width=0.4\textwidth,clip=true,keepaspectratio=true]{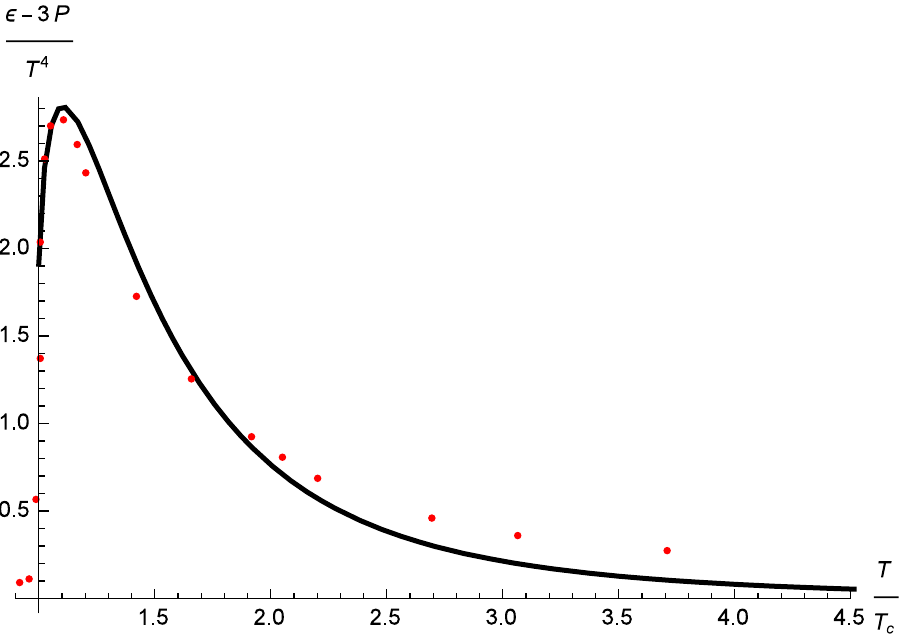}
  \end{center}
  \caption{The trace anomaly $\epsilon-3p$ from the quenched DhQCD model comparing with the SU(3) lattice data for pure gluon system from \cite{Boyd:1996bx}. The figure is taken from Ref.~\cite{Li:2014hja}, with $\mu_G=0.75\rm{GeV}$ and $G_5=1.25$.}
  \label{trace-anomaly-pure}
\end{figure}

In Fig.\ref{trace-anomaly-pure}, the result of the trace anomaly $\epsilon-3p$ of the hot gluonic matter is shown together with the SU(3) lattice data for pure gluon system \cite{Boyd:1996bx}. From the figure, it is easy to see that the results from the quenched dynamical holographic QCD model agree with the lattice data quite well. A sharp peak near $T_c$ appears in the trace anomaly, which implies that the correct IR physics is encoded in the 5D model.

Besides the thermodynamic quantities, the transport coefficients, which characterize the response of the hot system when deviating from equilibrium states, also connect the theoretical calculations with the experiments. Therefore, after checking the effectiveness of the DhQCD model in describing the thermodynamic quantities, we would like to extend the study to the transport properties, including the jet quenching parameter, the bulk viscosity as well as the shear viscosity. 

\begin{figure}
  \begin{center}
    \includegraphics[width=0.4\textwidth,clip=true,keepaspectratio=true]{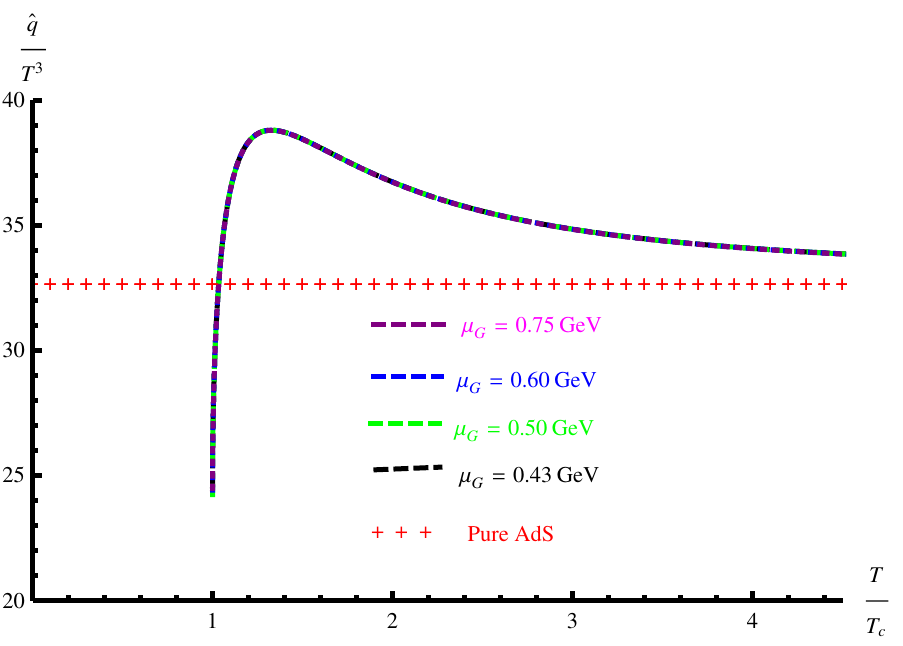}
  \end{center}
  \caption{The $T^3$ sclaed jet quenching parameter $\hat{q}/T^3$ as a function of temperature $T/T_c$, which is extracted from the quenched DhQCD model . The figure is taken from Ref.~\cite{Li:2014hja}, with $\mu_G=0.75\rm{GeV}$ and $G_5=1.25$.}
  \label{qhat-pure}
\end{figure}

When an energetic parton pass through a hot dense medium, due to the interaction with the medium, it would lose energy to the medium. The jet quenching parameter $\hat{q}$ is introduced to measure the energy loss rate. According to the dictionary proposed in Ref.\cite{Liu:2006ug},  $\hat{q}$ is related to the adjoint light like Wilson loop 
\begin{equation}
W^{Adj}[\mathcal {C}]\approx exp(-\frac{1}{4\sqrt{2}}\hat{q}L^{-}L^2).
\end{equation}
In the holographic model, the expectation value of the Wilson loop could be extracted from the on-shell string Nambu-Goto action of the dual string configuration with the 4D loop as its boundary. It is not difficult to derive the expression of $\hat{q}$ in our holographic model  \cite{Li:2014hja} as
\begin{equation}\label{qhatfor-res}
\hat{q}=\frac{\sqrt{2}\sqrt{\lambda}}{\pi z_h^3 \int_0^{1}d\nu\sqrt{\frac{e^{-4A_s(\nu z_h)}}{z_h^4}\frac{1-f(\nu z_h)}{2}f(\nu z_h)}}.
\end{equation}

We insert the background solution into the above equations, and get the numerical results for the jet quenching parameter, as shown in Fig.\ref{qhat-pure}. Different from the constant results in pure AdS background, the temperature scaled jet quenching parameter $\hat{q}/T^3$ in the quenched DhQCD model exhibit a non-trivial behavior, especially near the critical temperature $T_c$. Below $T_c$, $\hat{q}/T^3$ increases fast with temperature while it decreases smoothly above $T_c$. A peak appears at around $T=1.1 T_c$, with a height around 40. This behavior is quite similar to the peak of the trace anomaly. Both of these behaviors 
show the successful encoding of the novel property in the quenched DhQCD model.

\begin{figure}[h]
\begin{center}
\includegraphics[width=0.4\textwidth]{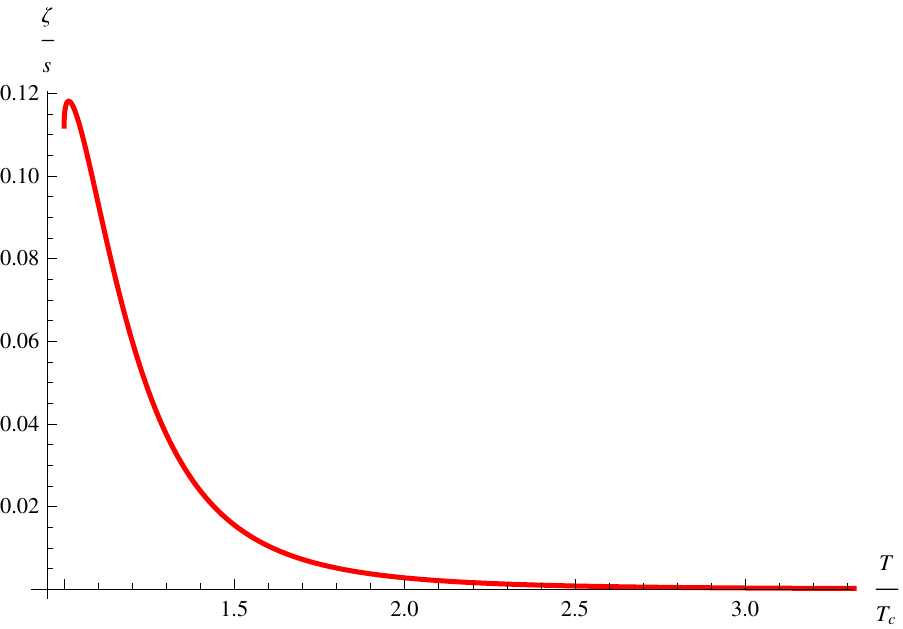} \hspace*{0.1cm}
\hskip 0.15 cm
\textbf{( a ) }\\
\includegraphics[width=0.4\textwidth]{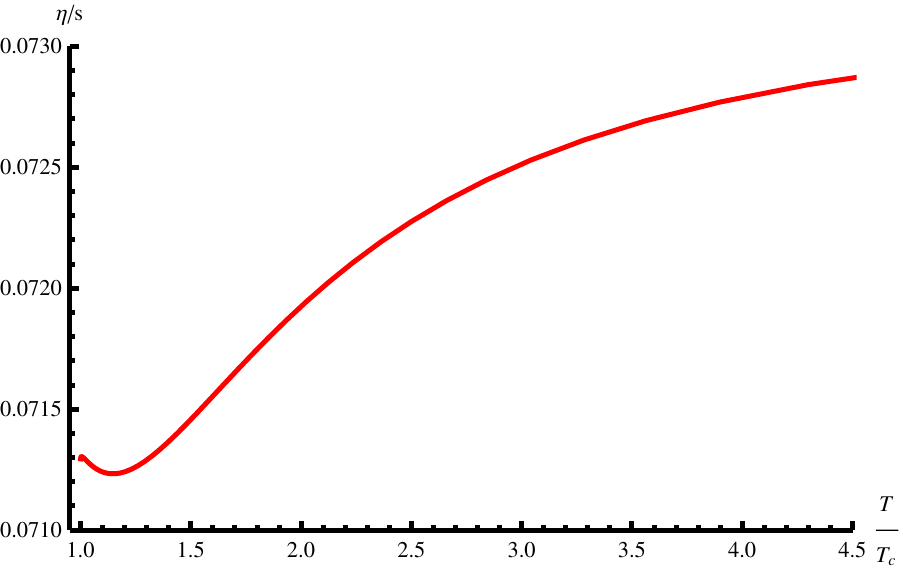}
 \hskip 0.15 cm 
 \textbf{( b )} \\
\end{center}
\caption[]{
(a) Bulk viscosity over entropy density $\zeta/s$ and (b) shear viscosity over entropy density $\eta/s$ as functions of the temperature, from the quenched DhQCD model. The figures are taken from Ref.~\cite{Li:2014dsa}. }
 \label{EOS-Transp}
\end{figure}

The viscosity coefficients are also important input in the hydrodynamic evolution. Theoretically, the bulk viscosity is related to the two point function of the stress tensor through the Kubo formula 
\begin{equation}
\zeta=\frac{1}{9}\underset{\omega\rightarrow0}{\text{lim}}\frac{1}{\omega}\text{Im}\langle T_{xx}(\omega)T_{xx}(0)\rangle.
\end{equation}
Here, $T_{xx}$ represents the diagonal components of the stress tensor. From the holographic dictionary, the two point function could be worked out by considering the perturbation of the diagonal components of the metric $\delta g_{xx}\equiv h_{xx}$. Then one can derive the equation of motion for the metric perturbation as 
\begin{eqnarray}\label{hxx-eom}
h_{xx}^{''}=&&(-\frac{1}{3A^{'}}-4A^{'}+3B^{'}-\frac{f^{'}}{f})h_{xx}^{'}\nonumber\\
&&+(-\frac{e^{-2A+2B}}{f^{2}}\omega^2+\frac{f^{'}}{6f A^{'}}-\frac{f^{'}B^{'}}{f})h_{xx},\nonumber\\
\end{eqnarray}
with $e^{2A}=e^{2A_s-\frac{4}{3}\Phi}$, $e^{2B}=\frac{e^{2A_s-\frac{4}{3}\Phi}}{Z^{'2}}$ and $z^{'}=Z(z)\equiv\sqrt{\frac{8}{3}}\mu_G^2z^2$.
After solving $h_{xx}$ numerically, one obtains 
\begin{eqnarray}
{\rm Im} G_R=-\frac{e^{4A-B}f}{16\pi G_5A^{'2}}|\text{Im} h_{xx}^{*}h_{xx}^{'}|,
\end{eqnarray}
and the numerical results for $\zeta$ could be obtained. In Fig.\ref{EOS-Transp}(a), the numerical results of bulk viscosity over entropy density $\zeta/s$ are given. Again, we could see that near $T_c$ a peak apears in the $\zeta/s$ line, which is qualitatively in agreement with the lattice results in Refs.~\cite{Kharzeev:2007wb,Karsch:2007jc,Meyer:2007dy}.

Similar with the bulk viscosity, the shear viscosity can also be worked out through the Kubo formula 
\begin{equation}
\eta=\underset{\omega\rightarrow0}{\text{lim}}\frac{1}{\omega}\text{Im}\langle T_{xy}(\omega)T_{xy}(0)\rangle.
\end{equation}
The only difference is that in the shear case one has to consider the off-diagonal components $T_{xy}$, and correspondingly the perturbation of $\delta g_{xy}\equiv h_{xy}$. Then one can derive the equation of motion for $h_{xy}$ and after solving it one gets the numerical results of $\eta$. Unfortunately, it is proved that for any isotropic Einstein gravity, the value of $\eta/s$ would always be $1/4\pi$, no matter how one deforms the background metric. Such a trivial constant $\eta/s$ result is different from many well known systems, like water, helium, nitrogen and so on, of which $\eta/s$ show a minimum near the transition temperature. To get a nontrivial behavior of $\eta/s$, following Ref.~\cite{Cremonini:2012ny}, we consider the higher derivative terms 
\begin{equation}
S_1=\frac{1}{16\pi G_5}\int d^5x \sqrt{-g}\big(\beta e^{\sqrt{2/3}\gamma\Phi}R_{\mu\nu\lambda\rho}R^{\mu\nu\lambda\rho}\big),
\end{equation}
as a probe, where $\beta,\gamma$ are two parameters controlling the contribution of the higher derivative term. Considering the metric perturbation in such a gravity system, one could derive $\eta/s$ up to the order of $O(\beta)$ \cite{Cremonini:2012ny} as 
\begin{eqnarray}
\frac{\eta}{s}=\frac{1}{4\pi}\left(1-\frac{\beta}{c_0}e^{\sqrt{2/3}\Phi_h}(1-\sqrt{2/3}\gamma z_h\Phi^{'}(z_h))\right),
\end{eqnarray}
with $c_0=-z_h^5\partial_z\left((1-z^2/z_h^2)^2e^{2A}/(8f(z)z^2)\right)|_{z=z_h}$. Then we take $\beta=0.01,\gamma=-\sqrt{8/3}$ and show the numerical results in Fig.\ref{EOS-Transp}(b). Interestingly, we see that a valley appear at $T=1.1T_c$. Therefore, from the holographic study, it indicates that the phase transition has non-trivial effect on the transport coefficients. For more details about this part, please refer to Ref.~\cite{Li:2014hja,Li:2014dsa}.


\subsection{Chiral phase transition}
\label{sec-phasetransition}
\begin{figure}[h]
\begin{center}
\includegraphics[width=0.4\textwidth]{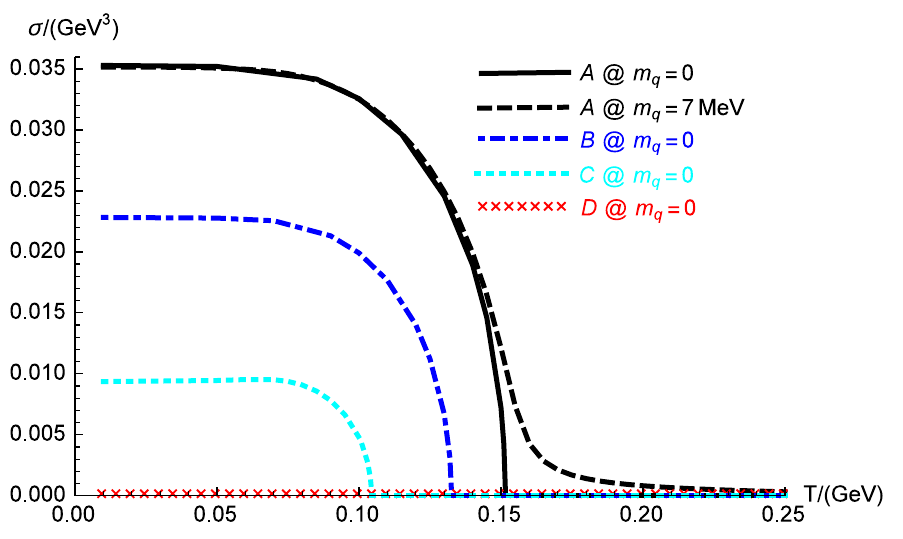} 
\hskip 0.15 cm \textbf{( a ) } \\
\includegraphics[width=0.4\textwidth]{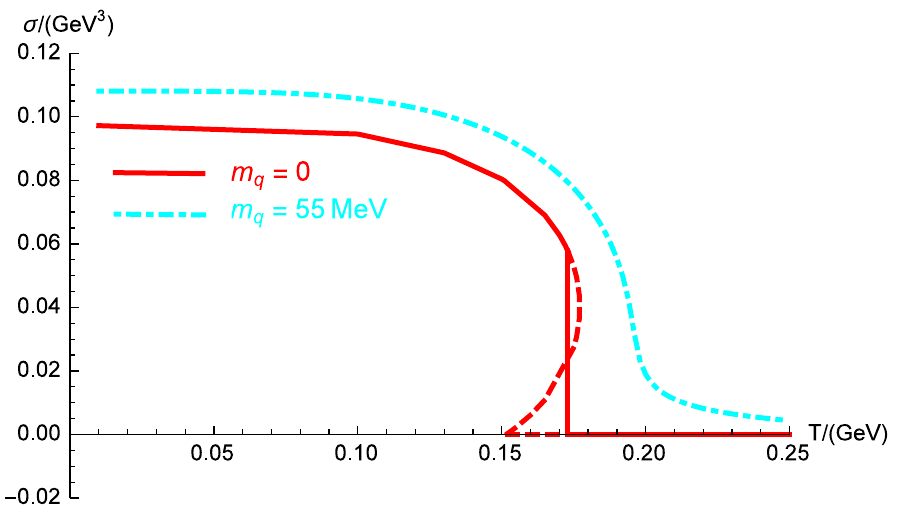}
\hskip 0.15 cm \textbf{( b )} \\
\end{center}
\caption[]{The chiral condensate $\sigma$ as a function of the temperature $T$ in 2-flavor case (a) and 3-flavor case (b),
respectively. The figures are taken from Refs.~\cite{Chelabi:2015cwn,Chelabi:2015gpc}.}
\label{chiral}
\end{figure}

In the previous sections, the dynamical holographic QCD model is shown to give a good description of glue-dynamics and thermodynamics. The deconfinement phase transition could be described as well.  In this subsection, we will transfer to the flavor dynamics and focus on the chiral phase transition in this model. Since solving the DhQCD model with finite temperature, density is not an easy task, we will treat the flavor part as a probe. Within such a scenario, the dynamical part of the action would reduce to the soft-wall model action
\begin{eqnarray}\label{action-soft-simple}
 S=-\int d^5x
 \sqrt{-g}e^{-\Phi}Tr(D_m X^+ D^m X+V_X(|X|).
\end{eqnarray}
Here, since only the scalar part is relevant for chiral phase transition, we neglect the other parts in the full action. As mentioned above, there is a  $SU(N_f)_L\times SU(N_f)_R$ gauge symmetry in this action. But if the scalar field $X$ has a non-zero value $X_0$, this symmetry would be broken to $SU(N_f)_V$. That is the holographic realization of the 4D spontaneously breaking of chiral symmetry. Considering the symmetry of the background, one has $X_0=\frac{\chi(z)}{\sqrt{2N_f}}I_{N_f}$, with $I_{N_f}$ the $N_f\times N_f$ identity matrix. Since chiral condensate is homogeneous at finite temperature, $\chi(z)$ depends on the holographic coordinate $z$ only. Inserting those conditions into the action, one reaches the effective form of the action
\begin{equation}\label{eff-action}
S_{\chi}=-\int d^5x
 \sqrt{-g}e^{-\Phi}(\frac{1}{2}g^{zz}\chi^{'2}+V(\chi)).
\end{equation}
We take the following simple scalar potential
\begin{equation}\label{profile-chi}
V(\chi)\equiv Tr({V_X(|X|)})=-\frac{3}{2}\chi^2+v_3 \chi^3+v_4 \chi^4,
\end{equation}
where we have taken the mass term as $-3X^+X$, the cubic term comes from the determinant of $X$ and the quartic term $v_4 \chi^4$ is introduced to get non-zero condensate at low temperature. Different from the 4D field theory, such a potential is insufficient to get non-zero condensate. It is shown that the derivative of the dilaton profile $\Phi^\prime(z)$ should take negative values in the scale of chiral dynamics. Combining with the requirement of the linear confinement in the confinement scale, we take the following form for the dilaton field
\begin{equation}\label{int-dilaton}
\Phi(z)=-\mu_1z^2+(\mu_1+\mu_0)z^2\tanh(\mu_2z^2),
\end{equation}
which is the interpolation of the requirement in the scales of chiral and confinement dynamics. Though there are other kinds of modification, but interestingly all of them requires two scales to incoorperate the two dynamics, which is in agreement with the instanton liquid model studied in \cite{Shuryak:1997vd}. To get the reasonable values of the Regge slope, the vacuum condensate and the transition temperature, we take the parameters as $\mu_0=(0.43)\rm(GeV)^2\simeq0.18\rm{GeV}^2, \mu_1=(0.83 {\rm GeV})^2\simeq0.69 {\rm GeV}^2$  and $\mu_2=(0.176 {\rm GeV})^2\simeq 0.03 {\rm GeV}^2,v_3=0, v_4=8$ for two flavor case. As for three flavor case, to consider the instanton effect, we replace the value of $v_3$ with $v_3=-3$. If all the quark masses are degenerate, the equation of motion could be derived as 
\begin{eqnarray}\label{eom-chi-1}
\chi^{''}+(3A_s^{'}-\Phi^{'}+\frac{f^{'}}{f})\chi^{'}- \frac{e^{2A_s}}{f}\partial_\chi V(\chi)=0.
\end{eqnarray}
The UV expansion of the solution could be obtained as $m_q \zeta z+\frac{\sigma}{\zeta} z^3$, with $m_q$ the quark mass,  $\sigma$ the chiral condensate, and $\zeta=\sqrt{N_c}/2\pi$\cite{Cherman:2008eh}. By requiring the regularity of the solution at the horizon, one can solve $\sigma$ as a function of $m_q$ and $T$. The results are given in Fig.\ref{chiral}(a) for two flavor with $v_3=0$ and in Fig.\ref{chiral}(b) for three flavor with $v_3=-3$.

From the figure, one could see that: in the chiral limit of two flavor case, the phase transition is a second order transition, and any finite quark mass would drive the transition to a crossover one; in the chiral limit of three flavor case, the phase transition turns to be a first order one, and only sufficient large quark mass($>37\rm{MeV}$) could drive it to a crossover transition.  

\begin{figure}
  \begin{center}
    \includegraphics[width=0.4\textwidth,clip=true,keepaspectratio=true]{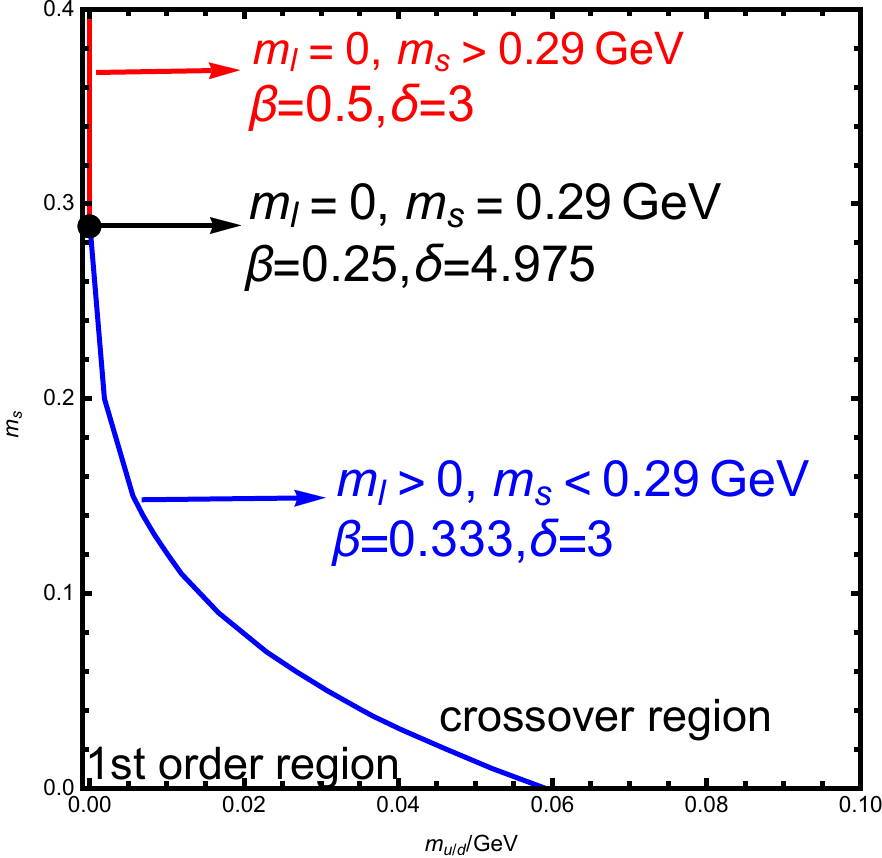}
  \end{center}
  \caption{The phase diagram of chiral phase transition in the quark mass plane. The figure is taken from Ref.~\cite{Chen:2018msc}.}
  \label{chiral-mass-diag}
\end{figure}

The above study could be extended to `2+1' flavors case, i.e. $m_u=m_d\neq m_s$. Then the diagonal part of the scalar field $X$ becomes different, we have $\chi_u=\chi_d\equiv\chi_l\neq\chi_s$. Then we have to solve the coupled equations
\begin{eqnarray}
\chi_l^{''}&+&(3A_s^{'}-\Phi^{'}+\frac{f^{'}}{f})\chi_l^{'}\nonumber\\
&+&\frac{e^{2A_s}}{f}(3\chi_l-3v_3\chi_l\chi_s-4v_4\chi_l^3)=0,\label{eom-chil}\\
\chi_s^{''}&+&(3A_s^{'}-\Phi^{'}+\frac{f^{'}}{f})\chi_s^{'}\nonumber\\
&+&\frac{e^{2A_s}}{f}(3\chi_s-3v_3\chi_l^2-4v_4\chi_s^3)=0.\label{eom-chis}
\end{eqnarray}
Similarly, the quark masses and condensates could be read from the UV expansion as $\chi_l=m_l \zeta z+...+\sigma_l/\zeta z^3+..., \chi_s=m_s \zeta z+...+\sigma_s/\zeta z^3+...$, with $m_l=m_u=m_d$ the light quark masses, $m_s$ the strange quark mass, and $\sigma_l=\langle\bar{u}u\rangle=\langle\bar{d}d\rangle, \sigma_s=\langle\bar{s}s\rangle$.
Then one can solve the condensates from the above equations and get the mass dependence behavior of the phase transition. The results are summarize in Fig.\ref{chiral-mass-diag}. It is shown that the whold mass plane is divided into two parts, which are separated by a critical line (the blue and red segments in Fig.\ref{chiral-mass-diag}). In the corner near the three flavor chiral limit, the chiral phase transition is a first order phase transition, while it is a continuous crossover in the other side of the critical line. It should also be noted that there is a tri-critical point in the critical line, which divides the critical line into two parts with different critical exponents, as shown in the figure.  Qualitatively, the results agrees quite well with the `Columbia plot'\cite{Brown:1990ev,Ding:2015ona}. For more details, please refer to Ref.~\cite{Chelabi:2015cwn,Chelabi:2015gpc,Li:2016smq,Chen:2018msc}.

\section{Discussion and summary}
\label{sec-summary}

We have reported our studies on hadron physics and QCD matter in the framework of DhQCD model, which is constructed
in the graviton-dilaton-scalar framework, where the dilaton background field and scalar field are dual to the gluon condensate and 
the chiral condensate operator thus can represent the gluodynamics (linear confinement) and chiral dynamics (chiral symmetry breaking), 
respectively. The dilaton background field and the scalar field are a function of the 5th dimension, which plays the role the energy scale, 
in this way, the DhQCD model can resemble the renormalization group from ultraviolet (UV) to infrared (IR). By solving the Einstein equation, 
the metric structure at IR is automatically deformed by the nonperturbative gluon condensation and chiral condensation in the vacuum.

It is seen that the pure gluon system can be well described by the graviton-dilaton system with a quadratic dilaton field $\Phi(z)=\mu z^2$, 
the deformed metric can be analytically solved. With this deformed metric, the glueball/oddball spectra are in good agreement with lattice result,
and the thermodynamical properties of this pure gluon system including the energy density, the pressure density and the trace anomaly agree with 
lattice data well, and the temperature dependent transport coefficients including the shear viscosty, bulk viscosity and jet quenching parameter carry 
the information of the phase transition, i.e, the ratio of the jet quenching parameter over cubic temperature ${\hat q}/T^3$,the ratio of the bulk viscosity
over the entropy density show a peak around the critical temperature $T_c$, and the ratio of the shear viscosity over the entropy density shows a 
valley around the critical temperature $T_c$, at which the scaled trace anomaly $(\epsilon-3 p)/T^4$ also shows a peak.

Then we add the flavor part as a probe on the pure gluon system, the full DhQCD, i.e., the deformed metric can be self-consistently solved including 
both the gluon condensate and chiral condensate. On this deformed metric we can solve the light-flavor meson spectra, which is in agreement with 
experimental data. The flavor part itself is also extended to four flavor case, and the obtained heavy meson spectra also agree well with experimental data.
Furthermore, the chiral phase transition is successfully realized in the framework of holographic QCD framework. 

In summary, our DhQCD model can describe hadron physics, QCD phase transition, thermodynamical properties and transport properties of QCD matter quite 
successfully. In this sense, we can use this DhQCD model as an effective method for non-perturbative QCD.

\begin{acknowledgements}
This work is supported in part by the National Natural Science Foundation of China (NSFC) Grant  Nos 11725523, 11735007, 11805084 and supported by the Strategic Priority Research Program of Chinese Academy of Sciences under Grant Nos XDB34030000 and XDPB15, the start-up funding from University of Chinese Academy of Sciences(UCAS), the Fundamental Research Funds for the Central Universities, the China Postdoctoral Science Foundation under Grant No. 2021M703169, and Guangdong Pearl River Talents Plan under Grant No. 2017GC010480.
\end{acknowledgements}

\bibliographystyle{unsrt}
\bibliography{main.bib}

\end{document}